\documentclass[12pt]{article}

\usepackage{lscape}

\usepackage{hhline}

\usepackage[dvips,final]{graphicx}
\usepackage{wrapfig}

\author{A.S.~Yurkov \\ 
Omsk, Russia, 644076, e-mail:fitec@mail.ru }

\title{ On the Continual Theory of Flexoelectric Deformations }

\begin{document}

\maketitle

\begin{abstract}

In many cases the correct theoretical description of  flexoelectricity requires the consideration of the finite size of a body and is reduced to the solution of  boundary  problems for partial differential equations. Generally speaking, in this case one should solve  jointly the equations of polarization equilibrium  and equations of elastic equilibrium. However, due to the fact that typically  flexoelectric moduli  are  very small, usually one can  consider the solution of  polarization  equilibrium equations at a given elastic strain (direct flexoelectric effect) or the solution of elastic equilibrium  equations at given polarization (converse flexoelectric effect). Derivation of the polarization equilibrium equations  and boundary conditions for them can be made in the quite usual way. Solution of these equations  usually is not too difficult problem. On the contrary description of converse flexoelectric effect is more complicated problem. Inter alia  effective solution of corresponded boundary problems  requires the development of special mathematical methods. The subject of this paper is a detailed discussion of  the relevant theory  focuses on the converse  flexoelectric effect. It is also considered some particular examples illustrating the application of the general theory.

\end{abstract}

\section{Introduction}

A body in which flexoelectrical phenomena are observed  always has a finite size. In some cases  the finite sample size is not significant, and the sample can be described as an infinite medium. Examples are the influence  of the flexoelectric effect on the dispersion of acoustic phonons \cite{bib:Axe70} or its influence on structure of ferroelectric domain wall   \cite{bib:DW}. In other cases, such as flexoelectric bending of a plate \cite{bib:Bursian68}, finite size of a sample plays a fundamental role.

Obviously, the effects associated with the finite size of the sample caused by the presence of the sample boundary. In particular, this raises the surface piezoelectricity which, as it turns out, leads to the effects of the same order as the flexoelectric effect \cite{bib:TagYu12}. 

Less trivial fact is  that  even without  surface piezoelectricity,  in the presence of flexoelectricity  classical theory of elasticity  should be modified.  In this case it turns out  that, generally speaking, one should  use elastic boundary conditions of non-classical form. These boundary conditions are consistent only if one considers the spatial dispersion of the elastic moduli (contribution to the thermodynamic potential which is a bilinear in elastic strain gradients) \cite{bib:Yurkov11}. The latter leads to that  not only the boundary conditions, but also differential equations  of elastic equilibrium should be non-classical.

Equations of elastic equilibrium are needed to describe the converse flexoelectric effect (mechanical response to the polarization). As regards the direct flexoelectric effect (polarization response to an inhomogeneous elastic strain), to describe it the equations of polarization equilibrium are needed. It is essential that this  equations  and boundary conditions for them are  modified by flexoelectricity not so radically as the equations and boundary conditions describing the elastic response (converse flexoelectric effect). This is related to the fact that the flexoelectric contributions to the thermodynamic potential contains only the first spatial derivatives of the polarization, while it contains the second spatial derivatives of the elastic displacements. Equations of polarization equilibrium and boundary conditions for them in the presence of flexoelectricity were derived in paper \cite{bib:Eliseev09}.

Thus,  non-trivial features of flexoelectricity  in finite samples  reveal themselves mainly in the converse flexoelectric effect. This is why  we  do not discuss much  equations of polarization equilibrium, focusing mainly on the converse flexoelectric effect at a given polarization. Moreover, in the particular examples (but not in the general theory) further we assume a homogeneous polarization at which non-trivial features of  flexoelectric effect in finite samples reveal itself  most dramatically.

The case of homogeneous polarization is interesting in the sense that a naive analysis based on constitutive equations leads to the (erroneous) conclusion that the homogeneous polarization does not cause a flexoelectric deformation of a body. The consequence of this conclusion is the  ability to create a sensor which not behave as an actuator \cite{bib:Cross06,bib:Chu09}. Such an ability contradicts the general principles of thermodynamics. But in reality violations of the principles of thermodynamics is not happening, homogeneous polarization does lead to a flexoelectric deformation of the body \cite{bib:TagYu12,bib:Yurkov11}.

In the paper \cite{bib:TagYu12} the conclusion that homogeneous polarization  leads to a flexoelectric deformation of a thin plate was made on the basis of the particular ansatz for elastic displacement distribution over the sample and direct minimization of the thermodynamic potential. A more rigorous analysis of this problem requires the solution of differential equations of elastic equilibrium with appropriate boundary conditions. Such a solution for the case of homogeneously polarized ball was presented in the paper \cite{bib:Yurkov13}. As was expected the solution confirmed the presence of flexoelectric strains caused by homogeneous polarization.

It is worth mentioning that in the presence of flexoelectricity the  exact (within the framework of a continuum media theory) equations of elastic equilibrium are too complex to be used for a case differs from a simple case of a ball. Even for  a ball the solution is very cumbersome and requires an introduction of non-standard functions presented by a series in powers of radial coordinate. So that the development of approximate method is worth. Such a method was developed in the paper \cite{bib:Yurkov14}.

Papers on the subject   \cite{bib:Yurkov13,bib:Yurkov14} are too brief. More detailed description is presented here. General aspects of derivation of the equations of elastic equilibrium and the boundary conditions for them in the case when the thermodynamic potential depends on higher derivatives discussed in another paper \cite{bib:YuBoundChap}. Here we apply these results to the case of flexoelectricity and develop the theory further. Besides, it is presented a comparison of exact solution \cite{bib:Yurkov13} and  the solution of the same problem within the approximate theory \cite{bib:Yurkov14}. Some additional examples of  approximate theory application are presented also. 

\section{Thermodynamic potential, differential equations of elastic equilibrium,  and boundary condition }

It is well known \cite{bib:YudinTag13} that within the theory of continuum flexoelectricity is described by  a contribution to thermodynamic potential density:
\begin{equation}
\label{eq:hflx1}
{\cal H}_{flx}= - f^{(1)}_{klij} P_k u_{i,j,l}- f^{(2)}_{klij}P_{k,l}u_{i,j} \, .
\end{equation}
Here $P_k$ is polarization, $u_i$ is elastic displacement, $f^{(1)}_{klij}$ and $f^{(2)}_{klij}$ are  material tensors, $(\dots)_{,i}=\partial(\dots)/\partial x_i\,$,  and Einstein  summation rule is adopted. Note that usually  strain tensor $u_{ij}=(u_{i,j}+u_{j,i})/2$ is used instead of displacement gradients $u_{i,j}$ (distortion tensor). But since $u_{i,j}$ is convolved with material tensors which are symmetrical in corresponding indices, it does not matter to write down $u_{i,j}$ or $u_{ij}\,$. Note also that here it is used definition of  $f^{(a)}_{klij}$ slightly different  from that adopted in \cite{bib:YudinTag13}, pairs of indices $ij$ and $kl$ are swapped. This definition corresponds to \cite{bib:Yurkov11,bib:Yurkov13,bib:Yurkov14}  and more widely accepted. Certainly  the opposite definition is also applicable. 

Thermodynamic potential of a body $H$  is integral over body volume $V$ from thermodynamic potential density. Using integration by parts its flexoelectric  contribution  can be represented in the form:
\begin{equation}
H_{flx}=  \frac{f_{klij}}{2}\int \left(P_{k,l} u_{i,j}  -  P_k u_{i,j,l}\right) dV + \oint d_{ijk}u_{i,j}P_k dS \, ,
\end{equation}
where $f_{klij}=f^{(1)}_{klij} - f^{(2)}_{klij}$ is so-called flexocoupling tensor, $d_{ijk}$ is surface piezoelectric tensor of special form:
\begin{equation}
\label{eq:dijk}
d_{ijk}=-\frac{1}{2}\left(f^{(1)}_{klij}+f^{(2)}_{klij}\right)n_l \, .
\end{equation}
Hereinafter $n_l$ is unit vector normal to body surface $S$. Note that $d_{ijk}$ is different in different points of the surface and it is symmetric in indices $i,j$.

Thus flexoelectric part of thermodynamic potential density (\ref{eq:hflx1}) can be reduced to the Lifshitz-type form 
\begin{equation}
\label{eq:hflxlt}
{\cal H}_{flx}=\frac{1}{2}f_{klij}\left(P_{k,l} u_{i,j}  -  P_k u_{i,j,l}\right)
\end{equation}
together with additional surface piezoelectricity described by piezoelectric tensor (\ref{eq:dijk}). Further effects of surface piezoelectricity are not described so that  Lifshitz-type form (\ref{eq:hflxlt}) is used.
 
Certainly besides flexoelectric potential (\ref{eq:hflxlt}) one should take in account elastic energy. It will be clear from what follows that besides conventional elastic energy $c_{ijkl}u_{i,j}u_{k,l}/2$ higher elasticity term $v_{ijklnm}u_{i,j,n}u_{k,l,m}/2$ should be added. So that   total volume density of the thermodynamic potential ${\cal H}_B$ is
\begin{equation}
\label{eq:hb}
{\cal H}_B=\frac{1}{2} \left[ c_{ijkl}u_{i,j}u_{k,l} +  v_{ijklnm}u_{i,j,n}u_{k,l,m} +
f_{klij}\left(P_{k,l} u_{i,j}  -  P_k u_{i,j,l}\right)\right] .
\end{equation}
Possible surface contribution to the thermodynamic potential is not considered here. So that volume integral from (\ref{eq:hb}) is total thermodynamic potential of a body.

Starting from the equation (\ref{eq:hb}) it is  possible to write down the equations of elastic equilibrium and the boundary conditions for them. In fact, one just need to calculate the tensors
\begin{equation}
\label{eq:tij}
T_{ij} = \frac{\partial {\cal H}_B}{\partial u_{i,j}} =
c_{ijkl}u_{k,l} + \frac{1}{2}f_{klij}P_{k,l}
\, ,
\end{equation}
\begin{equation}
\label{eq:thetaijk}
\Theta_{ijk} = \frac{\partial {\cal H}_B}{\partial u_{i,j,k}} =
v_{ijnlkm}u_{n,l,m} - \frac{1}{2}f_{nkij}P_n \, ,
\end{equation}
and use the results of \cite{bib:YuBoundChap}. In this way it turns out that ``physical stress'' $\sigma_{ij}$ which obey the equations of equilibrium 
\begin{equation}
\label{eq:equilib}
\sigma_{ij,j}=0
\end{equation}
are determined by equation:
\begin{equation}
\sigma_{ij}=T_{ij} - \Theta_{ijk,k}=c_{ijkl}u_{k,l} - v_{ijnlkm}u_{n,l,m,k} + f_{klij}P_{k,l} \, .
\end{equation}
Boundary conditions for (\ref{eq:equilib}) can be written in different equivalent forms, one should only substitute  (\ref{eq:tij}) and (\ref{eq:thetaijk}) to equations from \cite{bib:YuBoundChap}. Particularly they can be written in the form which coincides with that presented in \cite{bib:Yurkov11}:
\begin{equation}
\label{eq:thetnn}
\left.\Theta_{ijk}n_jn_k\right|_S=0 \, ,
\end{equation}
\begin{equation}
\left.\sigma_{ij}n_j - \Theta_{ijk,j}n_k + \Theta_{ijk,l}n_ln_jn_k -\Theta_{ijk}\gamma_{jk}
\right|_S=0 \, .
\end{equation}
Naturally terms contained $\Sigma_{ij}$ and $F_i$ do not appear  because surface contributions to thermodynamic potential is not described.

It is obvious from (\ref{eq:thetnn}) and (\ref{eq:thetaijk}) that if $v_{ijklnm} \to 0$ and $\left.f_{ijkl}P_i\right|_S \ne 0$ then generally $u_{i,j,k} \to \infty$. So that the limit $v_{ijklnm} \to 0$ is singular. This fact explains the above statement that generally in the presence of flexoectricity  one should  take into account the higher elasticity (spatial dispersion of elastic moduli). Higher elasticity term can omitted only in very special (and physically doubtful) case when $P_i=0$ on the surface of a body. In this special  case one can use classical boundary conditions, additional terms in them disappear.
  
Regarding the description of the direct flexoelectric effect, to obtain the corresponding differential equations and boundary conditions, one should  add  to (\ref{eq:hflxlt})  the following expression:
\begin{equation}
{\cal H}_P = \frac{1}{2}g_{ijkl}P_{i,j}P_{k,l} +\frac{1}{2}a_{ij}P_iP_j - E_iP_i \, ,
\end{equation}
where $E_i$ is electric field. If surface  contributions are not considered then  it remains only to vary such thermodynamic potential in $P_i$ and to obtain by standard way  the following differential equations 
\begin{equation}
a_{lk}P_l - g_{ijkl}P_{i,j,l} - f_{klij}u_{i,j,l} - E_k = 0 
\end{equation}
and boundary conditions for them
\begin{equation}
\left.
g_{ijkl}P_{i,j}n_l + \frac{1}{2}f_{klij}u_{i,j}n_l \right|_S = 0 \, . 
\end{equation}
It does not constitute a serious problem to take into account the surface contributions here. Usually such contributions contain only the polarization $P_i$ but not polarization  gradients $P_{i,j}\,$. This is why to find  the change of the equation written above  is a simple exercise in this case, in which we are not resting here.

\section{Exact solution for homogeneously polarized ball of isotropic dielectric}

When solving a particular problem it is convenient to use covariant formalism (see \cite{bib:YuBoundChap}). To transform equations from Cartesian form to covariant form one should only change partial derivatives to covariant derivatives and provide an index balance (only superscript can be convolved with subscript). So that equations of mechanical equilibrium and boundary conditions for them take the form:
\begin{equation}
\label{eq:el-eq-covar}
\sigma^{\alpha\beta}_{\phantom{\alpha\beta};\beta} = 0 \, ,
\end{equation}
\begin{equation}
\label{eq:elbond1covar}
\left.\Theta^{\alpha \beta \gamma}n_{\beta}n_{\gamma}\right|_S = 0 \, ,
\end{equation}
\begin{equation}
\label{elbond3}
\left.\sigma^{\alpha\gamma}n_{\gamma}  - 
\Theta^{\alpha\beta\gamma}_{\phantom{\alpha\beta\gamma };\beta}n_{\gamma} +
\Theta^{\alpha\beta\gamma}_{\phantom{\alpha\beta\gamma };\delta }
n^{\delta}n_{\beta}n_{\gamma} 
-\Theta^{\alpha\beta\gamma}\gamma_{\beta\gamma}\right|_S = 0 \, .
\end{equation}
Here $n_{\alpha}$ is covariant representation of unit vector normal to body surface $S$, tensors $\sigma^{\alpha\beta}$ and $\Theta^{\alpha\beta\gamma}$ are determined as follows:
\begin{equation}
\label{eq:sigma-def-covar}
\sigma^{\alpha\beta} = c^{\alpha\beta\gamma\delta}u_{\gamma;\delta} -
 v^{\alpha\beta\gamma\delta\varepsilon\zeta}u_{\gamma;\delta;\zeta;\varepsilon} 
+ f^{\gamma\delta\alpha\beta } P_{\gamma;\delta }  
 \, ,
\end{equation}
\begin{equation}
\Theta^{\alpha\beta\gamma}=
v^{\alpha\beta\varepsilon\delta\gamma\zeta}u_{\varepsilon;\delta;\zeta} -
\frac{1}{2}f^{\delta\gamma\alpha\beta } P_{\delta} \, .
\end{equation}
It is also convenient to use curvilinear coordinate system of special class in which equation of body surface has the form $x^3={\rm const}$. In this coordinate system tensor $\gamma^{\alpha\beta}$ can be presented in terms of Cristoffel symbols \cite{bib:YuBoundChap}:
\begin{equation}
\label{eq:gamma-ab-cris}
\gamma_{\beta\gamma} = \Gamma^{\varepsilon}_{\gamma\delta}n^{\delta}n_{\beta}n_{\varepsilon}
+\Gamma^{\varepsilon }_{\beta \delta} n^{\delta}n_{\varepsilon }n_{\gamma}
-\Gamma^{\delta}_{\gamma\beta}n_{\delta} 
 \,  .
\end{equation}
Besides  $n_{\alpha}$ has the only component:
\begin{equation}
n_3=\frac{1}{\sqrt{g^{33}}} \, .
\end{equation}

For a case of a ball it is naturally to use spherical coordinate system:
\begin{equation}
\label{eq:xyz}
\left\{
\begin{array}{l}
\displaystyle
x=r\sin\theta\cos\psi \, ,  \\ \\
\displaystyle
y=r\sin\theta\sin\psi \, , \\ \\ 
\displaystyle
z=r\cos\theta \, .
\end{array} \right.
\end{equation}
Here $x\,$,  $y$ and $z$   are Cartesian coordinates,  $\psi=x^1\,$,   $\theta=x^2\,$, and $r=x^3$ are curvilinear ones. Equation of the ball surface has the form $r = R\,$,  so that these curvilinear coordinates   belong to the  special class mentioned above. 

Instead of curvilinear  indices $1\, $, $2\, $, and $3\, $ further it is used also the indices $\psi\,$,  $\theta\,$,  and $r\,$ respectively. So the  letters $r\, $,  $\psi$ and $\theta$ are excluded from notation of ``running'' indexes. Particular Cartesian indices are further denoted  as $x$, $y$ and $z$.

Metric tensor components $g_{\alpha\beta}\,$, its determinant $g\,$,  and  Cristoffel symbols $\Gamma^{\alpha}_{\beta\gamma}$ can be derived  directly from (\ref{eq:xyz}):
\begin{equation}
\left\{
\begin{array}{l}
\displaystyle
g_{\psi\psi}=r^2\sin^2\theta \, , \quad
g_{\theta\theta}=r^2 \, , \quad
g_{r r}= 1 \, , \quad
g=r^4\sin^2\theta \, ,   \\ \\
\displaystyle
\Gamma^{\psi}_{\psi\theta} = \Gamma^{\psi}_{\theta\psi} =  \cot\theta \, , \quad
\Gamma^{\psi}_{\psi r}  = \Gamma^{\psi}_{r\psi}  =  r^{-1} \, , \quad
\Gamma^{\theta}_{\psi\psi} =  -\sin\theta\cos\theta \, ,  \\ \\ 
\displaystyle
\Gamma^{\theta}_{\theta r}  =  \Gamma^{\theta}_{r \theta} =  r^{-1} \, , \quad 
\Gamma^{r}_{\psi\psi} =  -r\sin^2\theta \, , \quad 
\Gamma^{r}_{\theta\theta} =  -r \, . 
\end{array}
\right.
\end{equation}
Other components are zero. Since tensor $g_{\alpha\beta}$ is diagonal, inverse tensor $g^{\alpha\beta}$ is diagonal also and corresponding components   are  simply equal to the inverse diagonal components of $g_{\alpha\beta}\,$. Unit vector normal to the surface has the only component $n_r=1$.

Further one should specify the form of the material tensors for  isotropic media. It is well-known that elastic tensor of such a  media is defined by two parameters, say its components  $c_{12}$ and $c_{44}\,$ (Voigt notation is used), and can be expressed as follows:
\begin{equation}
c_{ijkl}=c_{12}\delta_{ij}\delta_{kl} + c_{44}(\delta_{ik}\delta_{jl} + \delta_{il}\delta_{jk}) \, .
\end{equation}
It is possible to show that flexocoupling tensor of isotropic media can be expressed analogously:
\begin{equation}
f_{ijkl}=f_{12}\delta_{ij}\delta_{kl} + f_{44}(\delta_{ik}\delta_{jl} + \delta_{il}\delta_{jk}) \, .
\end{equation}
For six-rank tensor of higher elastic moduli of isotropic media further it is used a simplified expression
\begin{equation}
\begin{array}{l}
\displaystyle
v_{ijlknm}   =  
v_1(\delta_{ij}\delta_{lm}\delta_{nk} + \delta_{ij}\delta_{nl}\delta_{mk} +
\delta_{ij}\delta_{nm}\delta_{lk}  
 +\delta_{ik}\delta_{jn}\delta_{lm} + \\
\displaystyle  
 \delta_{il}\delta_{jn}\delta_{km} + \delta_{im}\delta_{jn}\delta_{kl}  + 
\delta_{in}\delta_{jk}\delta_{lm} + \delta_{in}\delta_{jl}\delta_{km} +
\delta_{in}\delta_{jm}\delta_{kl} )  + \\
\displaystyle 
 v_2(\delta_{ik}\delta_{jl}\delta_{nm} + \delta_{ik}\delta_{jm}\delta_{nl}  
+\delta_{il}\delta_{jk}\delta_{nm} 
+ \delta_{il}\delta_{jm}\delta_{nk} + 
  \delta_{im}\delta_{jk}\delta_{nl} + \\
  \displaystyle
  \delta_{im}\delta_{jl}\delta_{nk} ) \, ,
\end{array}
\end{equation}
where $v_1$ and $v_2$ are constant parameters. 

Above the expressions of material tensors are presented in Cartesian components. To convert them to curvilinear components one should only change Kronecker delta to metric tensor with corresponding location (up or bottom) of indices. 

If the ball is polarized along $z$-axis with polarization $P$ then polarization vector has two covariant components:
\begin{equation}
P_r=P\cos\theta \, ,
\end{equation}
\begin{equation}
P_{\theta}=-rP\sin\theta \, .
\end{equation}

The equations  above mathematically completely determine the problem under study. To solve this problem one must first separate the variables. Generally this is done by decomposition of the vector field $u_{\alpha}$ in a series of spherical vectors. Note that since the  covariant formalism is used here,  spherical vectors do not coincide with their usual form \cite{bib:Varshalovich}. However, the required form of these vectors can easily be obtained from conventional. Covariant spherical vectors can be defined as follows
\begin{equation}
\left\{
\begin{array}{l}
\displaystyle
Y_{lm|\theta}^{(1)} = \frac{r}{\sqrt{l(l+1)}}(Y_{lm})_{,\theta} \, , \quad
Y_{lm|\psi}^{(1)}     =  \frac{r}{\sqrt{l(l+1)}}\, imY_{lm} \, , \\ \\
\displaystyle
Y_{lm|\theta}^{(2)} = \frac{r}{\sqrt{l(l+1)}}\cdot\frac{-im}{\sin\theta} Y_{lm} \, , \quad 
Y_{lm|\psi}^{(2)}     =  \frac{r}{\sqrt{l(l+1)}}\, \sin\theta (Y_{lm})_{,\theta} \, , \\ \\
\displaystyle
Y_{lm|r}^{(3)}    = Y_{lm} \, , \\ \\
\displaystyle
Y_{lm|r}^{(1)}    = 0 \, , \quad
Y_{lm|r}^{(2)}    = 0 \, , \quad
Y_{lm|\theta}^{(3)} = 0 \, , \quad
Y_{lm|\psi}^{(3)}     =  0 \, , \quad
\end{array}
\right.
\end{equation}
where $Y_{lm}$ is scalar spherical functions expressed in the usual manner in terms of  associated Legendre polynomials.

In the case considered here, i.e. when the polarization is homogeneous, expansion in spherical vectors is simplified radically. Indeed, it is easy to see that in such  expansion of the polarization field $P_{\alpha}$ there are  only the terms with $l=1\,$,  $m=0\,$ and top indices 1,3.  Since the theory is linear, it follows that the expansion of the field $u_{\alpha}$ contains only  the similar terms. As a result, this leads to the fact that  $u_r \sim \cos\theta\,$, $u_{\theta} \sim \sin\theta\,$, $u_{\psi}=0\,$. This is why one can use  a substitution:
\begin{equation}
\label{eq:upsi}
u_{\psi}=0 \, ,
\end{equation}
\begin{equation}
\label{eq:utheta}
u_{\theta}=-rf_2(r)\sin\theta
\end{equation}
\begin{equation}
\label{eq:ur}
u_r = f_1(r)\cos\theta \, .
\end{equation}
Factor $r$ added into  (\ref{eq:utheta}) to make  radial function $f_2(r)$ analytical at \mbox{$r=0\,$}. The consequence of using the covariant formalism is that $u_{\theta}$ is not a physical displacement in the meridional direction, such physical displacement is $u_{\theta}\sqrt{g^{\theta\theta}}$, while $g^{\theta\theta}=r^{-2}\,$. Certainly the physical component of the displacement should not have a singularity at the ball center. Note that  covariant form of $Y^{(1)}_{lm|\theta}$ contains the same factor $r$.

Thus, the problem is reduced to finding the radial functions $f_1(r)$ and $f_2(r)\,$. To find the corresponding differential equations and boundary conditions one should substitute (\ref{eq:upsi})--(\ref{eq:ur}) in the above equations and make a very cumbersome but absolutely straightforward  algebraic transformations. As the result it yields the system of two differential equations
\begin{equation}
\label{eq:feq1}
\begin{array}{l}
\displaystyle
v_3(\xi^4 f''''_1+4\xi^3 f'''_1-8\xi^2 f''_1+16 f_1)+v_3(8\xi^2 f''_2-16f_2) + \\ \\
\displaystyle
v_4(\xi^4 f''''_1+4\xi^3 f'''_1-6\xi^2 f''_1+12f_1) - \\ \\
 \displaystyle
v_4(2\xi^3 f'''_2-2\xi^2 f''_2-4\xi f'_2+12f_2) = \\ \\
 \displaystyle
c_{44}\xi^2(2 \xi^2 f''_1+4\xi f'_1-6f_1)-c_{44}\xi^2 (2 \xi f'_2-6 f_2) + \\ \\
 \displaystyle
 c_{12}\xi^2(\xi^2 f''_1+2\xi f'_1-2f_1)-c_{12}\xi^2(2\xi f'_2-2f_2)   \, ,
\end{array}
\end{equation}
\vspace{0.25cm}
\begin{equation}
\label{eq:feq2}
\begin{array}{l}
\displaystyle
v_3(\xi^4 f''''_2 + 4\xi^3 f'''_2 - 4\xi^2 f''_2 + 8f_2) +v_3( 4\xi^2f''_1 - 8f_1) - \\ \\
\displaystyle
v_4(2\xi^2 f''_2 - 4f_2) +v_4( \xi^3f'''_1 + 4\xi^2 f''_1 - 2\xi f'_1 - 4f_1) = \\ \\
\displaystyle
 c_{44}\xi^2(\xi^2 f''_2 + 2\xi f'_2  - 4f_2) +c_{44}\xi^2(\xi f'_1+ 4f_1) - \\ \\
\displaystyle
c_{12}\xi^2 2f_2 +c_{12}\xi^2(  \xi f'_1 + 2f_1)  \, ,
\end{array}
\end{equation}
\vspace{0.25cm}
and four boundary condition for it
\vspace{0.5cm}
\begin{equation}
\begin{array}{l}
\label{eq:fbc1}
\displaystyle
 2v_1( f'''_2 +  f''_2 - 18 f'_2 + 34 f_2 
+ 8 f''_1 + 14 f'_1 - 34 f_1) + \\ \\
\displaystyle
+ 2v_2(2 f'''_2 + 2 f''_2 - 20 f'_2 + 36 f_2 
+ 4 f''_1 + 16 f'_1 - 36 f_1)  -  \\ \\
\displaystyle
\left. - 2c_{44}R^2 ( f'_2 - f_2 + f_1)\right|_{\xi=1} =  f_{12}R^2 P   \, , 
\end{array}
\end{equation}
\vspace{0.25cm}
\begin{equation}
\label{eq:fbc2}
\begin{array}{l}
\displaystyle
R^2 [c_{12}( 2 f_1 - 2 f_2)  + (c_{12} + 2c_{44})  f'_1  ]  - \\ \\
\displaystyle
v_1(9f'''_1 + 18 f''_1 - 46  f'_1 + 36 f_1   -  
14  f''_2 + 32  f'_2 - 36 f_2) - \\ \\
\displaystyle
\left.2 v_2(3f'''_1 + 6 f''_1 - 22  f'_1 + 28 f_1 -  2  f''_2  + 16  f'_2 - 28f_2)\right|_{\xi=1}
= R^2 f_{12}P
  \, ,
\end{array}
\end{equation}
\vspace{0.25cm}
 \begin{equation}
 \label{eq:fbc3}
 \begin{array}{l}
 \left.(18 v_1 + 12 v_2)  f''_1 +
 12v_1 (3  f'_1 - 4 f_1 - 2  f'_2 + 4 f_2)\right|_{\xi=1}= \\ \\
 \displaystyle
  (f_{12} + 2 f_{44}) R^2  P \, ,
 \end{array}
 \end{equation}
 \vspace{0.25cm}
 \begin{equation}
 \label{eq:fbc4}
 \begin{array}{l}
 \displaystyle
 2 v_1( f''_2 + 2 f'_2 - 6f_2  + 2 f'_1 + 6f_1) + \\ \\
 \displaystyle
 \left.4 v_2 ( f''_2 - 2  f'_2 + 2 f_2 + 2  f'_1 - 2 f_1)\right|_{\xi=1} 
=  f_{44}R^2  P \, .
  \end{array}
 \end{equation}
Here $\xi=r/R$ is dimensionless radial coordinate, prime denotes the derivative in $\xi$, $v_3=(v_1+2v_2)R^{-2}\,$, $v_4=(8v_1+4v_2)R^{-2}\,$.

In general, the system of two fourth-order equations (\ref {eq:feq1}) --- (\ref {eq:feq2}) requires   eight  boundary conditions, but here we have  only four  boundary conditions (\ref{eq:fbc1}) --- (\ref{eq:fbc4}). However, as is  usually in the case of such problems, the missing boundary conditions are replaced by the conditions that a solution should be analytic at zero. Therefore, the solution should be expressed in the form of series in non-negative powers of $\xi$. 

Practically it is more convenient  to find the  complete basis set of solutions expressed as such series, and  to present a general solution as a linear combination of such basis functions. If the  number of  basis functions  is equal to the number of boundary conditions  then the coefficients in the linear combination can be  easily founded by solving a system of linear algebraic equations.

In accordance with the above, we express a solution in the form:
\begin{equation}
\label{eq:series}
f_i(\xi) = \sum_{k=1}^4 C_k{\cal B}_{ki}(\xi) \, ,
\end{equation}
\begin{equation}
{\cal B}_{ki}(\xi) = \sum_{n=0}^{\infty}a_{kin}\xi^n \, .
\end{equation}
The constant coefficients $a_{kin}$ are found from the condition that $ {\cal B}_{ki} (\xi) $ obeys the system of differential equations (\ref{eq:feq1}) - (\ref{eq:feq2}). This condition leads to the following linear algebraic equations for the coefficients (index $k$ which is numbering the solutions is omitted):
\begin{equation}
\label{eq:aeq}
\left\{
\begin{array}{l}
\displaystyle
[v_3(n^4-2n^3-9n^2+10n+16)  +  \\
\displaystyle
v_4(n^4-2n^3-7n^2+8n+12)]a_{1\,n}   \\
\displaystyle
+[v_3(8n^2-8n-16)  -v_4(2n^3 - 8n^2+2n+12)] a_{2\, n}  = \\
\displaystyle
[c_{44}(2n^2-6n-2) + c_{12}(n^2-3n)] a_{1 \, n-2}  - \\
\displaystyle
 [c_{44}(2n-10)+c_{12}(2n-6)]a_{2 \, n-2} \, , \\ \\
\displaystyle
[v_3(4n^2-4n-8)+ v_4(n^3+n^2-4n  -  4)]a_{1\,n} + \\
\displaystyle
[v_3(n^4-2n^3-5n^2+6n+8) - v_4(2n^2-2n-4)]a_{2\,n} = \\
\displaystyle
[c_{44}(n+2)  + c_{12}n]a_{1\,n-2}+[c_{44}(n^2-3n-2) - c_{12}2]a_{2\,n-2} \, .
\end{array}
\right.
\end{equation}

From (\ref{eq:aeq})  one can see that the system of linear algebraic equations, even though it is infinite, has the characteristic  structure: the coefficients with greater  $n$ are expressed in terms of the coefficients with smaller $n$. Therefore,  one can find $a_ {in}$ for any finite $n \,$ step by step, one needs only to analyse the case of  small $n$ and define  some of the $a_ {in}$ for these small $n \, $. 

Naturally for small $n$ system of equations (\ref{eq:aeq}) is degenerate, some $a_{in}$ with  small $n$ can be defined arbitrarily. This arbitrariness leads to that it may be several functions ${\cal B}_{ki}(\xi)$ and  their normalization may be arbitrary. Detailed analysis shows that the system of equations (\ref{eq:aeq}) is degenerate for $n=1,2,4$ only. It turns out five arbitrary constants, and so it is the same number of basis functions. However, one of this functions is physically meaningless and should be omitted (more on this see below).

From an abstract point of view, arbitrary constants can be defined as anything, it is just needed that  obtained basis functions be linearly independent. However, there is an important question: is it possible to choose these constants in such manner  that the series representing at least some of the basic functions cut short, would constitute a finite sums? It turns out that this choice of constants is really possible.

From the system of equations (\ref{eq:aeq})  it is clear that in order to series cut short,  it should be $a_{3i}=a_{4i}=0\,$. For this to be possible, in any case, right-hand sides of  (\ref{eq:aeq}) should be zero for $n=3,4$. Together with (\ref{eq:aeq})  for $n = 0,1,2$ these conditions  lead to a system of linear algebraic equations which is quite amenable to analysis. The result is that there may be two basic functions which are finite sums. 

One of them is determined by the conditions $a_{10}=a_{20}={\rm const} \ne 0\,$, $a_{n>0\,i}=0\,$. This function corresponds to the displacement of the ball as a whole. Such a displacement is physically meaningless and this basis function should be omitted. Certainly  such a displacement does not change the thermodynamic potential, so that this function  indeed is a solution of the equations derived from a stationary condition of this potential. But it is meaningless solution.
 
Another basis function, which is finite sum,  is determined by condition $(c_{12}-c_{44})a_{22}=(3c_{44}+2c_{12})a_{12}\,$, other coefficients are zero. So that first basis function is determined as follows:
\begin{equation}
\left\{
\begin{array}{l}
\displaystyle
{\cal B}_1: \\
\displaystyle
a_{12}=1\, ,\quad a_{22}=\frac{3c_{44} + 2c_{12}}{c_{12}-c_{44}} a_{12}\, , \\
\end{array}
\right.
\end{equation}
other  $a_{in}$  are  zero .

It should be stressed  that the function ${\cal B}_1$ obeys not only the equations (\ref{eq:feq1}) ---  (\ref{eq:feq2}) but also the equations of the classical theory of elasticity, which can be  obtained from (\ref{eq:feq1}) ---  (\ref{eq:feq2}) by setting $v_3=v_4=0\,$. It can be verified by direct substitution. 
 
Further analysis shows that there are three additional basic functions \mbox{${\cal B}_2$ --- ${\cal B}_4$} which are expressed by infinite series and can be chosen, for  example, as follows.

\begin{equation}
\left\{
\begin{array}{l}
\displaystyle
{\cal B}_2: \\
\displaystyle
a_{12}=1\, ,\quad a_{22}=\frac{3c_{44} + 2c_{12}}{c_{44}-c_{12}}a_{12} \, , \\
\displaystyle
a_{24}=0 \, ,\\
\displaystyle
 (40v_3 + 60 v_4)a_{14}  =  \\
 \displaystyle
 =(6c_{44}+ 4c_{12})a_{12} + (2c_{44}-2c_{12})a_{22} \, , \\
 \displaystyle
a_{i0}=a_{i1}=a_{i3}=0 \,  
\end{array}
\right.
\end{equation}
coefficients with $n\ge 5$ are calculated by means of (\ref{eq:aeq}).

\begin{equation}
\left\{
\begin{array}{l}
\displaystyle
{\cal B}_3: \\
\displaystyle
a_{11}=1\, ,\quad a_{21}=\frac{4v_3+3v_4}{4v_3+2v_4}a_{11} \, , \\
\displaystyle
a_{i0}=a_{i2}=a_{i4}=0 \, ,
\end{array}
\right.
\end{equation}
coefficients with $n=3$ and $n\ge 5$ are calculated by means of (\ref{eq:aeq}).

\begin{equation}
\left\{
\begin{array}{l}
\displaystyle
{\cal B}_4: \\
\displaystyle
a_{14}=1\, ,\quad a_{24}=\frac{2v_3+3v_4}{v_4-4v_3} a_{14} \, , \\
\displaystyle
a_{i0}=a_{i1}=a_{i2}=a_{i3}=0 \, ,
\end{array}
\right.
\end{equation}
coefficients with $n\ge 5$ are calculated by means of (\ref{eq:aeq}).

\begin{figure}[h]
\begin{center}
\includegraphics[width=13cm,keepaspectratio]{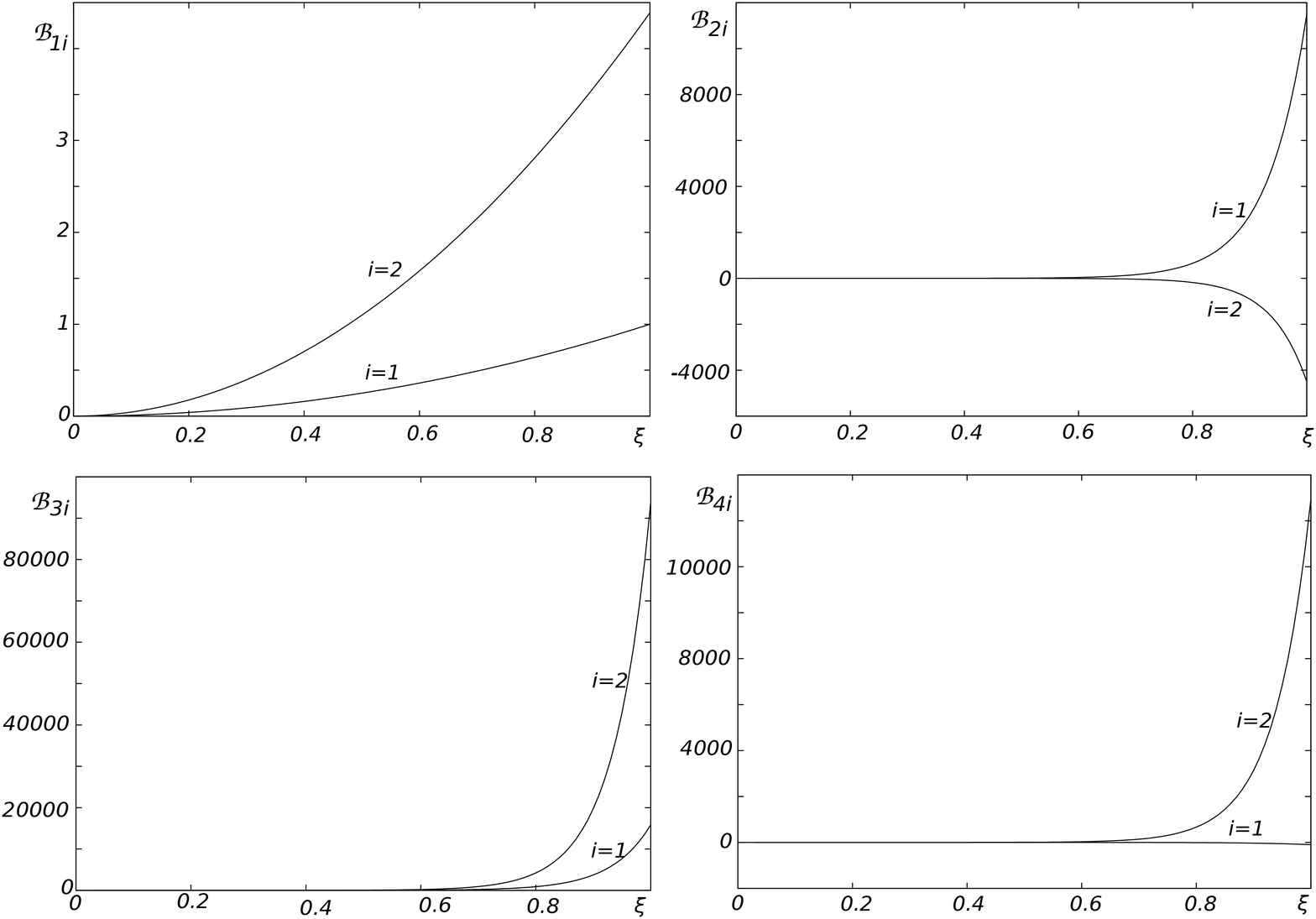}
\end{center}
%
\caption{Basis functions  for $R=1\cdot 10^{-5}\,$, $P=1\,$,  $c_{44}=1.1\cdot 10^{12}\,$, $c_{12}=3.4\cdot 10^{12}\, $, $f_{44}=f_{12}=1\cdot 10^{-3}\, $, $v_1=0.2\,$, $v_2=0.1\,$. }
\label{fig:B1-4m}
\end{figure}

Graphs of the  basis functions ${\cal B}_{ik}$ defined above for some set of parameters are shown in Fig.\ref{fig:B1-4m}. Turn our attention that in contrast with  ${\cal B}_{1i}$ the functions ${\cal B}_{2i}$ --- ${\cal B}_{4i}$ are concentrated near the surface and decay rapidly in the ball volume.

\begin{figure}[h]
\begin{center}
\includegraphics[width=10cm,keepaspectratio]{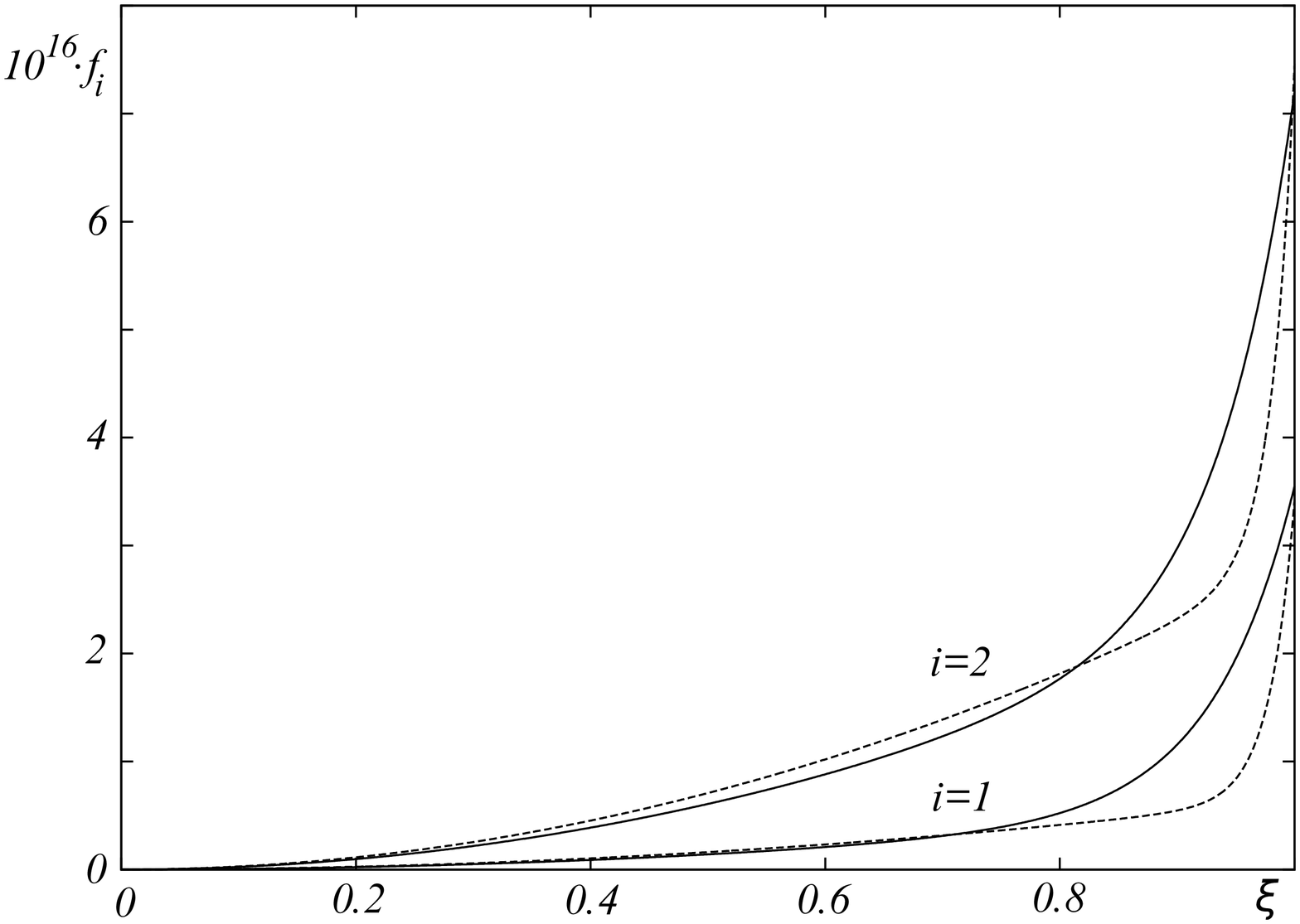}
\end{center}
%
\caption{Result of the calculation for $R=1\cdot 10^{-5}\,$, $P=1\,$,  $c_{44}=1.1\cdot 10^{12}\,$, $c_{12}=3.4\cdot 10^{12}\, $, $f_{44}=f_{12}=1\cdot 10^{-3}\, $. Solid line is for $v_1=2 \cdot 10^{-1}\,$, $v_2=1\cdot 10^{-1}\,$, dashed line is for $v_1=2\cdot 10^{-2}\,$, $v_2=1\cdot 10^{-2}\,$. }
\label{fig:exaball}
\end{figure}

Above there is explicit representation of four basic functions ${\cal B}_1$ --- ${\cal B}_4\,$.  It only remains to substitute (\ref{eq:series}) in the boundary conditions (\ref{eq:fbc1}) --- (\ref{eq:fbc4}), and from the resulting system of four linear algebraic equations to find numerically $C_k\,$. Thus, the problem is solved. Result of numerical calculations for some parameters is presented in the Fig.\ref{fig:exaball}.

Note that the graphs shows that, except for a  thin layer  near $\xi=1\, $, the curves are parabolic, the more so the  less  $v_1$ and $v_2\,$ are. In the bulk of the ball the only function  ${\cal B}_{1i}(\xi)\,$ remains with a good approximation.  Thus,  for a small $v_i\,$, except for a thin surface layer,  solution  approximately obeys the equations of classical elasticity theory (see  discussion of function ${\cal B}_1$ properties above). This observation is used in the next section to construct an approximate method of flexoelectric deformations calculations. This method is needed to describe more complicated geometry then spherical geometry of a ball.

\section{Approximate method to calculate flexoelectric deformations of finite size bodies}\label{sec:aprox}

Exact (within the framework of a continuum media theory) equations of elastic equilibrium  used in previous section  are  partial differential equations of the fourth order. They are too complex to be used in the cases interesting for applications. Even for a homogeneously polarized ball of isotropic media solution of them is cumbersome and requires a functions expressed by a series in powers of radial coordinate. So that to find flexoelectric deformation in more complicated cases  one needs an approximate method. Such a  method was proposed in \cite{bib:Yurkov14} and it is described in this section.

When solving the problem for a ball (see previous section) we have seen that the solution is a sum of two parts. The first part  corresponded basis function ${\cal B}_1\,$ obeys  the equations of the classical theory of elasticity, the second (contribution of remaining basis functions)  is a non-classical part  which, for small $v^{\alpha\beta\gamma\delta\varepsilon\zeta} \, $, is concentrated near the surface and rapidly decays  inside the body.  Hereafter   the first part is called   a volume  or a classical part, and the second one --- a non-classical or a surface part. 

It is naturally to assume that this property  should be preserved in  more general cases. The aim is to use this observation for  construction of approximation that is useful for solution of more general problems than the case of a ball.

By means of  (\ref{eq:sigma-def-covar})  equations (\ref{eq:el-eq-covar}) can be represent in following form
\begin{equation}
\label{eq:eqvil0}
c^{\alpha\beta\gamma\delta}u_{\gamma;\delta;\beta} +
f^{\gamma\delta\alpha\beta } P_{\gamma;\delta;\beta } -
 v^{\alpha\beta\gamma\delta\varepsilon\zeta}u_{\gamma;\delta;\zeta;\varepsilon;\beta} = 0 \, .
\end{equation}
Boundary conditions for this differential equations here are used in the form different from used in previous section. Particularly they are
\begin{equation}
\label{eq:bond1-0}
\Theta^{\alpha 3 3 } = 0 \, ,
\end{equation}
\begin{equation}
\label{eq:bond2-0}
\sigma^{\alpha 3} - \Theta^{\alpha \beta 3}_{\phantom{\alpha \beta 3}  ; \beta} + 
\Theta^{\alpha 3 3}_{\phantom{\alpha 3 3} , 3} +
\Theta^{\alpha (\beta\gamma)}\Gamma^3_{(\beta\gamma)} = 0 \, .
\end{equation}
Here the indices enclosed in parentheses   runs only the values 1 and 2,  if there are multiple indexes  in one pair of  parentheses then these indices are not equal to 3  simultaneously. Equivalence of this form of boundary conditions to (\ref{elbond3}) is proved in   \cite{bib:YuBoundChap}. Certainly it is assumed coordinate system where equation of the body surface has a form $x^3=x^3_S\,$,   $x^3_S$ is some constant.

According to the observation mentioned above   the solution of the  equation (\ref{eq:eqvil0}) is presented   in the form  $u_{\gamma}=\tilde{u}_{\gamma}+\hat{u}_{\gamma}\,$,  where $\tilde{u}_{\gamma}$ obey the classical equations:
\begin{equation}
\label{eq:tildeeq}
c^{\alpha\beta\gamma\delta}\tilde{u}_{\gamma;\delta;\beta} +
f^{\gamma\delta\alpha\beta } P_{\gamma;\delta;\beta } = 0 \, .
\end{equation}
From (\ref{eq:tildeeq}) and (\ref{eq:eqvil0}) it follows  that $\hat{u}_{\gamma}\,$ obey differential equations:
\begin{equation}
\label{eq:hateq}
c^{\alpha\beta\gamma\delta}\hat{u}_{\gamma;\delta;\beta}  -
v^{\alpha\beta\gamma\delta\varepsilon\zeta}\hat{u}_{\gamma;\delta;\zeta;\varepsilon;\beta} =
v^{\alpha\beta\gamma\delta\varepsilon\zeta}\tilde{u}_{\gamma;\delta;\zeta;\varepsilon;\beta} 
   \,  .
\end{equation}
Substitution  $u_{\gamma}=\tilde{u}_{\gamma}+\hat{u}_{\gamma}\,$ and some transformations also  yield that the boundary conditions (\ref{eq:bond1-0})  and (\ref{eq:bond2-0}) take the form:
\begin{equation}
\label{eq:bond1m}
v^{\alpha 3\varepsilon\delta 3 \zeta}\hat{u}_{\varepsilon;\delta;\zeta} =
\frac{1}{2}f^{\delta 3 \alpha 3} P_{\delta} - 
v^{\alpha 3\varepsilon\delta 3 \zeta}\tilde{u}_{\varepsilon;\delta;\zeta} \, ,
\end{equation}
\vspace{0.3cm}
\begin{equation}
\label{eq:bond2m}
\begin{array}{l} 
\displaystyle
c^{\alpha 3\gamma\delta}\tilde{u}_{\gamma;\delta} +
f^{ \gamma\delta \alpha 3  } P_{\gamma;\delta } -
 v^{\alpha 3\gamma\delta\varepsilon\zeta}\tilde{u}_{\gamma;\delta;\zeta;\varepsilon} - 
 v^{\alpha (\beta)\varepsilon\delta 3 \zeta}\tilde{u}_{\varepsilon;\delta;\zeta ; (\beta)} + \\ \\
\displaystyle
\frac{1}{2}f^{ \delta 3 \alpha (\beta) } P_{\delta ; (\beta)}  + 
v^{\alpha(\beta)\varepsilon\delta(\gamma)\zeta}\tilde{u}_{\varepsilon;\delta;\zeta}
\Gamma^3_{(\beta)(\gamma)} - \\ \\
\displaystyle
\frac{1}{2}f^{\delta (\gamma) \alpha(\beta)  } P_{\delta} \Gamma^3_{(\beta)(\gamma)} + 
c^{\alpha 3\gamma\delta}\hat{u}_{\gamma;\delta} -
 v^{\alpha 3\gamma\delta\varepsilon\zeta}\hat{u}_{\gamma;\delta;\zeta;\varepsilon} - \\ \\
\displaystyle
  v^{\alpha (\beta)\varepsilon\delta 3 \zeta}\hat{u}_{\varepsilon;\delta;\zeta ; (\beta)} + 
v^{\alpha(\beta)\varepsilon\delta(\gamma)\zeta}\hat{u}_{\varepsilon;\delta;\zeta}
\Gamma^3_{(\beta)(\gamma)}  = 0 \, .
\end{array}
\end{equation}
\vspace{0.5cm}

Equations above are exact. In order to turn into approximation one should note that  real values of  $v^{\alpha\beta\gamma\delta\varepsilon\zeta} \, $ are  small. It is convenient to assume that all of them are proportional to a scalar  $v \to 0 \,$. It is assumed also that all $c^{\alpha\beta\gamma\delta}\,$  is proportional to a scalar $c\,$.  If $v \to 0\,$ then  the right-hand side of (\ref{eq:hateq}) and the second term on the right hand side of (\ref{eq:bond1m}) can be neglected.  Indeed, since $\tilde{u}_{\gamma}$ obey the classical equation (\ref{eq:tildeeq}),  in the case of sufficiently smooth surface and in the absence of polarization gradients tend to infinity in the limit $v \to 0\,$ , the classical part can not have large gradients which can compensate smallness of $v^{\alpha\beta\gamma\delta\varepsilon\zeta} \, $. 
As for the non-classical part $\hat{u}_{\gamma}\,$, the situation is different: here such compensation is possible, but only if the derivatives are in   $x^3\,$. Moreover,  in a thin layer near the surface  covariant derivatives in $x^3\,$  can be replaced by the usual  derivatives,  $\Gamma$-terms give  only small corrections  here. Thus, (\ref{eq:hateq}) and  (\ref{eq:bond1m}) can be approximately replaced by
\begin{equation}
\label{eq:hateqm}
c^{\alpha 3 \gamma 3}\hat{u}_{\gamma , 3 , 3}  -
v^{\alpha 3 \gamma 3 3 3}\hat{u}_{\gamma , 3 , 3 , 3 , 3 } = 0 \, ,
\end{equation}
\begin{equation}
\label{eq:bond1m2}
v^{\alpha 3\varepsilon 3 3 3}\hat{u}_{\varepsilon , 3 , 3 } =
\frac{1}{2}f^{\delta 3 \alpha 3} P_{\delta} \, .
\end{equation}

Naturally decomposition  $u_{\gamma}=\tilde{u}_{\gamma}+\hat{u}_{\gamma}\,$  is not unique. By requiring that $\tilde{u}_{\gamma}$ obey (\ref{eq:tildeeq}), this  non-uniqueness is  restricted, but is not eliminated  completely. It is clear that $\tilde{u}_{\gamma} $ does not contain a non-classical part, but it does not mean that $\hat{u}_{\gamma} $ does not contain a classical part. To eliminate remaining non-uniqueness  one should require that $\hat{u}_{\gamma}$ exponentially decay  inside the body. Under such condition  $\hat{u}_{\gamma}$ obey not only  (\ref{eq:hateqm}) but also
\begin{equation}
\label{eq:hateqmint}
c^{\alpha 3 \gamma 3}\hat{u}_{\gamma}  -
v^{\alpha 3 \gamma 3 3 3}\hat{u}_{\gamma , 3 , 3 } = 0 \, .
\end{equation}

Equations (\ref{eq:hateqmint}) are a system of three ordinal differential equations, the dependence on $x^1$ and $x^2$ is parametric here. Moreover,  in a thin layer near the body surface one can assume  that the coefficients  do not depend on $x^3\,$, they are approximately equal  to the surface values. Solution of such a system is easy. In a standard way one should find a fundamental basis set of solutions in the form $\hat{u}_{\gamma}=\bar{u}_{\gamma}e^{\lambda(x^3-x^3_S)}\,$. Obviously equations for amplitudes $\bar{u}_{\gamma}$  are
\begin{equation}
\label{eq:eigen}
c^{\alpha 3 \gamma 3}\bar{u}_{\gamma } =
\lambda^2 v^{\alpha 3 \gamma 3 3 3}\bar{u}_{\gamma} 
 \,  .
\end{equation}

Equation (\ref{eq:eigen}) is a standard generalized eigenvalue problem for  symmetric positive definite $ 3\times 3$ matrices.  Therefore  $\lambda_n$ and  $\bar{u}_{\gamma}^n$ are calculable by standard way. If the body bulk corresponds to $x^3 \leq x^3_S\,$ for definiteness, then $\lambda_n$ equals the positive square root  of the  $n$-th eigenvalue, and general representation of non-classical part is
\begin{equation}
\label{eq:hatuapprox}
\hat{u}_{\gamma} = \sum_{n=1}^3 a_n \bar{u}^n_{\gamma}e^{\lambda_n(x^3-x^3_S)}\, ,
\end{equation}
where only the coefficients $a_n$ are  unknown. The latter can be found easily by (\ref{eq:bond1m2}) which yields a simple system of linear algebraic equations:
\begin{equation}
\label{eq:anfind}
\sum_{n=1}^3 \lambda_n^2 v^{\alpha 3\gamma 3 3  3}\bar{u}_{\gamma}^n a_n =
\frac{1}{2}f^{\delta 3 \alpha 3} P_{\delta} \, .
\end{equation}

Note that   the smallness of $v^{\alpha 3 \gamma 3 3 3}$  in  (\ref{eq:anfind})  is compensated   by $\lambda_n^2\,$,  and using (\ref{eq:eigen}) this system of equations   can be rewritten  as follows:
\begin{equation}
\sum_{n=1}^3  c^{\alpha 3\gamma 3}\bar{u}_{\gamma}^n a_n =
\frac{1}{2}f^{\delta 3 \alpha 3} P_{\delta} \, .
\end{equation}
Thus, the non-classical part of elastic displacements is found completely in explicit form.

As it follows from above,  to find  completely the non-classical part of elastic displacements one needs only the boundary conditions  (\ref{eq:bond1m}).  The remaining boundary conditions (\ref{eq:bond2m}) yield  the boundary conditions for the classical equations (\ref{eq:tildeeq}) in this  way, one should only substitute known $\hat{u}_{\gamma}$ to (\ref{eq:bond2m}) and select the terms  of the appropriate order of accuracy.

However, within this  derivation of the boundary conditions for  the equations  (\ref{eq:tildeeq}) a singularity arises requiring additional analysis. This singularity is related to the fact that it appears   a subtraction of two close but large terms in which it is impossible to use the approximate expression for $\hat{u}_{\gamma}$ derived above. In these (and only these) terms it is necessary to consider a corrections to such $\hat{u}_{\gamma}\,$. The appropriate analysis is made below  and while $\hat{u}_{\gamma}$  means the exact non-classical part of elastic displacements rather then approximate one defined by (\ref{eq:hatuapprox}).

According to the above,  all the terms in (\ref{eq:bond2m}), where gradients of $\tilde{u}_{\gamma}$ are convolved with $v^{\alpha\beta\gamma\delta\varepsilon\zeta}\,$,  should be omitted for the same reasons as in the case of the equation (\ref{eq:hateq}).  Further it is convenient to introduce the notation:
\begin{equation}
\label{eq:sdef}
\begin{array}{l}
\displaystyle
s^{\alpha} =
 - c^{\alpha 3\gamma\delta}\hat{u}_{\gamma;\delta} +
 v^{\alpha 3\gamma\delta\varepsilon\zeta}\hat{u}_{\gamma;\delta;\zeta;\varepsilon} +
  v^{\alpha (\beta) \gamma \delta 3 \zeta}\hat{u}_{\gamma;\delta;\zeta ; (\beta)} - \\ \\
\displaystyle   
v^{\alpha(\beta)\varepsilon\delta(\gamma)\zeta}\hat{u}_{\varepsilon;\delta;\zeta}
\Gamma^3_{(\beta)(\gamma)}   \, .
\end{array}
\end{equation}
With this notation the boundary conditions to the classical equations  can be written as follows:
\begin{equation}
\label{eq:bond3m2}
c^{\alpha 3\gamma\delta}\tilde{u}_{\gamma;\delta} =  s^{\alpha} - 
 f^{\gamma\delta \alpha 3  } P_{\gamma;\delta } - 
 \frac{1}{2}f^{ \delta 3 \alpha (\beta) } P_{\delta ; (\beta)} +  
\frac{1}{2}f^{\delta (\gamma) \alpha(\beta) } P_{\delta} \Gamma^3_{(\beta)(\gamma)}
  \, .
\end{equation}

Now all the terms to be further simplified are contained in $s^{\alpha}\,$. Carrying out  simplification,  first of all one should leave in the last term only the partial (not  covariant) derivatives in $x^3\,$. The reason for this is the same as above. Next, one should  express the remaining covariant derivatives in terms of  partial derivatives and $\Gamma$-terms. The exact expression for the third covariant derivative of the vector is very cumbersome. But keeping in mind that these third covariant derivatives are convolved with $v^{\alpha\beta\gamma\delta\varepsilon\zeta}\,$, one can  omit all the terms  where  initial vector is not differentiated at least  two times. So   it turns out:
\begin{equation}
\begin{array}{l}
\displaystyle
\hat{u}_{\alpha ; \beta ; \gamma ; \zeta} \approx 
\hat{u}_{\alpha , \beta , \gamma , \zeta}
-  \hat{u}_{ \delta , \beta , \zeta}\Gamma^{\delta}_{\alpha\gamma} 
-  \hat{u}_{ \delta , \gamma , \zeta}\Gamma^{\delta}_{\alpha \beta}
-  \hat{u}_{ \delta , \beta , \gamma}\Gamma^{\delta}_{\alpha\zeta} - \\ \\
\displaystyle
  \hat{u}_{\alpha , \delta , \zeta}\Gamma^{\delta}_{\beta\gamma} - 
  \hat{u}_{\alpha , \delta , \gamma}\Gamma^{\delta}_{\beta\zeta}
-  \hat{u}_{\alpha , \beta ,\delta }\Gamma^{\delta}_{\gamma \zeta}
\, .
\end{array}
\end{equation}
With this equation and simplifications  mentioned above we get:
\begin{equation}
\label{eq:sdefm2}
\begin{array}{l}
\displaystyle
s^{\alpha} = 
 - c^{\alpha 3\gamma\delta}\hat{u}_{\gamma;\delta} 
+ v^{\alpha 3\gamma\delta\varepsilon\zeta} \hat{u}_{\gamma , \delta , \zeta , \varepsilon} 
-   v^{\alpha 3\gamma\delta\varepsilon\zeta} \hat{u}_{ \rho , \delta , \varepsilon}\Gamma^{\rho}_{\gamma\zeta} - \\ \\
 \displaystyle
 v^{\alpha 3\gamma\delta\varepsilon\zeta} \hat{u}_{ \rho , \zeta , \varepsilon}
\Gamma^{\rho}_{\gamma \delta} 
- v^{\alpha 3\gamma\delta\varepsilon\zeta}\hat{u}_{ \rho , \delta , \zeta}
\Gamma^{\rho}_{\gamma \varepsilon}
- v^{\alpha 3\gamma\delta\varepsilon\zeta} \hat{u}_{\gamma , \rho , \varepsilon}
\Gamma^{\rho}_{\delta \zeta} - \\ \\
\displaystyle
 v^{\alpha 3\gamma\delta\varepsilon\zeta} \hat{u}_{\gamma , \rho ,\zeta}
\Gamma^{\rho}_{\delta\varepsilon}
- v^{\alpha 3\gamma\delta\varepsilon\zeta} \hat{u}_{\gamma , \delta ,\rho }
\Gamma^{\rho}_{\zeta \varepsilon} 
+ v^{\alpha (\beta) \gamma \delta 3 \zeta}\hat{u}_{\gamma ,\delta ,\zeta , (\beta)} - \\ \\
\displaystyle
 v^{\alpha (\beta) \gamma \delta 3 \zeta}\hat{u}_{ \rho , \delta , (\beta)}
 \Gamma^{\rho}_{\gamma\zeta} 
- v^{\alpha (\beta) \gamma \delta 3 \zeta} \hat{u}_{ \rho , \zeta , (\beta)}
\Gamma^{\rho}_{\gamma \delta}  - 
 v^{\alpha (\beta) \gamma \delta 3 \zeta} \hat{u}_{ \rho , \delta , \zeta}
\Gamma^{\rho}_{\gamma (\beta)} - \\ \\
\displaystyle
 v^{\alpha (\beta) \gamma \delta 3 \zeta} \hat{u}_{\gamma , \rho , (\beta)}
\Gamma^{\rho}_{\delta \zeta}  - 
 v^{\alpha (\beta) \gamma \delta 3 \zeta} \hat{u}_{\gamma , \rho , \zeta}
\Gamma^{\rho}_{\delta (\beta)} 
-  v^{\alpha (\beta) \gamma \delta 3 \zeta} \hat{u}_{\gamma , \delta ,\rho }
\Gamma^{\rho}_{\zeta (\beta)} - \\ \\
\displaystyle
 v^{\alpha(\beta)\varepsilon 3 (\gamma) 3}\hat{u}_{\varepsilon ,3 , 3}
\Gamma^3_{(\beta)(\gamma)}   \, .
\end{array}
\end{equation}
Again, using the fact that to compensate  smallness of  $v^{\alpha\beta\gamma\delta\varepsilon\zeta}\,$ one needs  at least  two  derivatives in $x^3\,$,  this equation is further simplified:
\begin{equation}
\label{eq:sdefm3}
\begin{array}{l}
\displaystyle
s^{\alpha} = 
 - c^{\alpha 3\gamma\delta}\hat{u}_{\gamma;\delta} 
+ v^{\alpha 3\gamma\delta\varepsilon\zeta} \hat{u}_{\gamma , \delta , \zeta , \varepsilon} 
-   v^{\alpha 3\gamma 3 3 \zeta} \hat{u}_{ \rho ,  3 , 3 }\Gamma^{\rho}_{\gamma\zeta}  - \\ \\
\displaystyle
 v^{\alpha 3\gamma\delta  3 3 } \hat{u}_{ \rho ,  3  , 3 }\Gamma^{\rho}_{\gamma \delta} 
- v^{\alpha 3\gamma 3 \varepsilon 3}\hat{u}_{ \rho , 3  , 3}\Gamma^{\rho}_{\gamma \varepsilon}
- v^{\alpha 3\gamma\delta 3\zeta} \hat{u}_{\gamma ,  3  , 3 }\Gamma^{3}_{\delta \zeta} - \\ \\
\displaystyle
 v^{\alpha 3\gamma\delta\varepsilon 3} \hat{u}_{\gamma , 3  , 3}\Gamma^{3}_{\delta\varepsilon}
- v^{\alpha 3\gamma 3 \varepsilon\zeta} \hat{u}_{\gamma ,  3 , 3 }\Gamma^{3}_{\zeta \varepsilon} 
+ v^{\alpha (\beta) \gamma 3 3 3}\hat{u}_{\gamma , 3 , 3 , (\beta)} - \\ \\
\displaystyle
  v^{\alpha (\beta) \gamma  3 3  3} \hat{u}_{ \rho ,  3 , 3}\Gamma^{\rho}_{\gamma (\beta)}
- v^{\alpha (\beta) \gamma \delta 3 3} \hat{u}_{\gamma , 3 , 3}\Gamma^{3}_{\delta (\beta)} - \\ \\
\displaystyle
v^{\alpha (\beta) \gamma 3 3 \zeta} \hat{u}_{\gamma , 3  , 3 }\Gamma^{3}_{\zeta (\beta)}
- v^{\alpha(\beta)\varepsilon 3 (\gamma) 3}\hat{u}_{\varepsilon ,3 , 3}
\Gamma^3_{(\beta)(\gamma)}   \, .
\end{array}
\end{equation}
It is convenient to introduce yet another  notation:
\begin{equation}
\label{eq:hdef}
\begin{array}{l}
\displaystyle
h^{\alpha\rho}
= v^{\alpha 3\gamma 3 3 \zeta} \Gamma^{\rho}_{\gamma\zeta} 
+ v^{\alpha 3\gamma\delta  3 3 }\Gamma^{\rho}_{\gamma \delta} 
+ v^{\alpha 3\gamma 3 \varepsilon 3}\Gamma^{\rho}_{\gamma \varepsilon}
+ v^{\alpha 3\rho\delta 3\zeta} \Gamma^{3}_{\delta \zeta} + \\ \\
\displaystyle
 v^{\alpha 3\rho\delta\varepsilon 3}\Gamma^{3}_{\delta\varepsilon}
+ v^{\alpha 3\rho 3 \varepsilon\zeta}\Gamma^{3}_{\zeta \varepsilon} 
+  v^{\alpha (\beta) \gamma  3 3  3} \Gamma^{\rho}_{\gamma (\beta)} + \\ \\
\displaystyle
v^{\alpha (\beta) \rho \delta 3 3} \Gamma^{3}_{\delta (\beta)}
+  v^{\alpha (\beta) \rho 3 3 \zeta} \Gamma^{3}_{\zeta (\beta)}
+ v^{\alpha(\beta)\rho 3 (\gamma) 3} \Gamma^3_{(\beta)(\gamma)}   \, .
\end{array}
\end{equation}
Using the symmetry properties of $ v^{\alpha \beta \gamma \delta \varepsilon \zeta} \, $ it can be rewritten as follows:
\begin{equation}
\label{eq:hdefmm}
\begin{array}{l}
\displaystyle
h^{\alpha\beta} =
 ( v^{\alpha 3\gamma 3 \delta 3}
+ 2v^{\alpha 3\gamma\delta  3 3 })\Gamma^{\beta}_{\gamma \delta} 
+ (2v^{\alpha 3\beta \gamma\delta 3}
+ v^{\alpha 3\beta\gamma 3\delta} )\Gamma^{3}_{\gamma \delta} + \\ \\
\displaystyle  
  2v^{\alpha 3 \beta \gamma  (\delta) 3}\Gamma^{3}_{\gamma (\delta)}  
+  v^{\alpha  3 \gamma  3 (\delta)  3} \Gamma^{\beta}_{\gamma (\delta)}
+ v^{\alpha (\gamma) \beta 3 (\delta) 3 } \Gamma^3_{(\gamma)(\delta)}   \, .
\end{array}
\end{equation}
In this notation the expression for $s^{\alpha}$  can be written as follows:
\begin{equation}
\label{eq:sdefm4}
\begin{array}{l}
\displaystyle
s^{\alpha} =  - c^{\alpha 3\gamma\delta}\hat{u}_{\gamma , \delta} + 
c^{\alpha 3\gamma\delta}\hat{u}_{\varepsilon}
\Gamma^{\varepsilon}_{\gamma\delta}
+ v^{\alpha 3\gamma\delta\varepsilon\zeta} \hat{u}_{\gamma , \delta , \zeta , \varepsilon} 
+ \\ \\
\displaystyle
v^{\alpha (\beta) \gamma 3 3 3}\hat{u}_{\gamma , 3 , 3 , (\beta)} - 
 h^{\alpha\beta}\hat{u}_{\beta , 3 , 3}  \, .
\end{array}
\end{equation}

Now one can proceed  to the analysis of the singularity mentioned above. It appears in the first and third term of (\ref{eq:sdefm4}), when all the derivatives  are in $x^3\,$. For all other terms there is no singularity and one can use the expression  (\ref{eq:hatuapprox}) for $\hat{u}_{\gamma}\,$. It is also clear that singular  terms exactly cancel each other out because if approximate expression (\ref{eq:hatuapprox}) is used. This is due to the explicit form of this expression.  One might think that these  terms should simply be excluded, but in reality the situation is more complicated. 

The fact is that    
$- c^{\alpha 3\gamma 3}\hat{u}_{\gamma , 3 , 3}  +  v^{\alpha 3\gamma 3 3 3}
\hat{u}_{\gamma , 3 , 3 , 3 ,3}$ is equal to zero only for  the main approximation for $\hat{u}_{\gamma}\,$. But the corrections to this  approximation may result in the fact that this expression becomes finite. Henceforth we denote  such corrections   as $w_{\gamma}$ keeping notation $\hat{u}_{\gamma}$  only for the main approximation. In this notation it turns out:
\begin{equation}
\label{eq:sdefm5}
\begin{array}{l}
\displaystyle
s^{\alpha} =  
- c^{\alpha 3\gamma 3}w_{\gamma , 3 }  +  v^{\alpha 3\gamma 3 3 3}
w_{\gamma , 3 , 3 , 3 }
- c^{\alpha 3\gamma (\delta)}\hat{u}_{\gamma , (\delta)} + 
c^{\alpha 3\gamma\delta}\hat{u}_{\varepsilon}
\Gamma^{\varepsilon}_{\gamma\delta} + \\ \\
\displaystyle
 v^{\alpha 3\gamma\delta\varepsilon\zeta} \hat{u}_{\gamma ( , \delta , \zeta , \varepsilon)} 
+ v^{\alpha (\beta) \gamma 3 3 3}\hat{u}_{\gamma , 3 , 3 , (\beta)} 
- h^{\alpha\beta}\hat{u}_{\beta , 3 , 3}  \, .
\end{array}
\end{equation}
To compensate  the smallness of $ v^{\alpha \beta \gamma \delta \varepsilon \zeta} \, $  in the fifth term of this equation there should be  at least two differentiations in $x^3$. Thus, after some reductions  one can rewrite (\ref{eq:sdefm5}) as  follows:
\begin{equation}
\label{eq:sdefm8}
\begin{array}{l}
\displaystyle
s^{\alpha} =   
- c^{\alpha 3\gamma 3}w_{\gamma , 3 }  +  v^{\alpha 3\gamma 3 3 3}
w_{\gamma , 3 , 3 , 3 }
- c^{\alpha 3\gamma (\delta)}\hat{u}_{\gamma , (\delta)} + 
c^{\alpha 3\gamma\delta}\hat{u}_{\varepsilon}
\Gamma^{\varepsilon}_{\gamma\delta} +  \\ \\
\displaystyle
 2( v^{\alpha 3\gamma (\beta) 3 3 } +  
v^{\alpha 3  \gamma 3 (\beta) 3})\hat{u}_{\gamma , 3 , 3 , (\beta)} 
- h^{\alpha\beta}\hat{u}_{\beta , 3 , 3}  \, .
\end{array}
\end{equation}

It is clear that in (\ref{eq:sdefm8}) one should  use only corrections $w_{\gamma}$ of order $v^{1/2}\,$. Only such corrections yield a contribution of the same order as the other terms. To find the corresponding contributions, we should keep in the exact equation 
\begin{equation}
\label{eq:hatuweq}
c^{\alpha\beta\gamma\delta}(\hat{u}_{\gamma}+
w_{\gamma})_{;\delta;\beta}  -
v^{\alpha\beta\gamma\delta\varepsilon\zeta}
(\hat{u}_{\gamma}+w_{\gamma})_{;\delta;\zeta;\varepsilon;\beta} =
v^{\alpha\beta\gamma\delta\varepsilon\zeta}\tilde{u}_{\gamma;\delta;\zeta;\varepsilon;\beta} 
   \,  
\end{equation}
only the terms of the order  $v^{-1/2}\,$. Indeed if $w_{\gamma} \sim v^{1/2}\,$ then the main terms $w_{\gamma , 3 , 3}$ and $v^{\alpha 3\gamma 3 3 3}w_{\gamma , 3 , 3 , 3 , 3}$ have just  such an order. Note that although there are terms $\hat{u}_{\gamma , 3 , 3}$ and $v^{\alpha 3\gamma 3 3 3}\hat{u}_{\gamma , 3 , 3 , 3 , 3}$ in the equation, which are of the order $\sim v^{-1}\,$, now they   cancel each other completely due to the fact that  now $\hat{u}_{\gamma}$ denotes the main approximation. Right hand side of (\ref{eq:hatuweq}) is $\sim v\,$, so  it should be omitted. The terms $c^{\alpha\beta\gamma\delta}w_{\gamma (;\delta;\beta)}\,$ and  $v^{\alpha\beta\gamma\delta\varepsilon\zeta}w_{\gamma (;\delta ;\zeta ;\varepsilon ;\beta)}\,$ should be omitted by the same reasons.  Thus (\ref{eq:hatuweq}) is reduced to
\begin{equation}
\label{eq:hatuwsimpl}
- c^{\alpha 3\gamma 3}w_{\gamma , 3 , 3}  +
v^{\alpha 3\gamma 3 3 3}w_{\gamma , 3 , 3 , 3 , 3} =
c^{\alpha\beta\gamma\delta}\hat{u}_{\gamma (;\delta ;\beta)}  -
v^{\alpha\beta\gamma\delta\varepsilon\zeta}
\hat{u}_{\gamma (;\delta ;\zeta ;\varepsilon ;\beta)}
   \,  .
\end{equation}

Equation (\ref{eq:hatuwsimpl}) requires further simplification: in the right-hand side we should keep only the terms  $\sim v^{-1/2}\,$. For this in the first term there should be one partial derivative in $x^3\,$, and in the second term there should be three derivatives in $x^3\,$. This is why we simplify second and fourth covariant derivatives as follows:
\begin{equation}
\hat{u}_{\gamma ; \delta ; \beta} \approx
\hat{u}_{\gamma , \delta , \beta} -
\hat{u}_{\gamma , \rho}\Gamma^{\rho}_{\delta\beta} -
\hat{u}_{\rho , \delta}\Gamma^{\rho}_{\gamma\beta} -
\hat{u}_{\rho , \beta}\Gamma^{\rho}_{\gamma\delta} \, ,
\end{equation}

\begin{equation}
\begin{array}{l}
\displaystyle
\hat{u}_{\gamma ; \delta ; \zeta ; \varepsilon ; \beta} \approx 
\hat{u}_{\gamma , \delta , \zeta , \varepsilon , \beta} -
\hat{u}_{\rho , \delta , \zeta , \varepsilon} \Gamma^{\rho}_{\gamma\beta} -
\hat{u}_{\gamma , \rho , \zeta , \varepsilon} \Gamma^{\rho}_{\delta\beta} - \\ \\
\displaystyle
 \hat{u}_{\gamma , \delta , \rho , \varepsilon} \Gamma^{\rho}_{\zeta\beta} -
\hat{u}_{\gamma , \delta , \zeta , \rho} \Gamma^{\rho}_{\varepsilon\beta} -
\hat{u}_{ \rho , \delta , \varepsilon , \beta }\Gamma^{\rho}_{\gamma\zeta} 
-  \hat{u}_{ \rho , \zeta , \varepsilon , \beta}\Gamma^{\rho}_{\gamma \delta} - \\ \\
\displaystyle
\hat{u}_{ \rho , \delta , \zeta , \beta}\Gamma^{\rho}_{\gamma\varepsilon}
-  \hat{u}_{\gamma , \rho , \varepsilon , \beta}\Gamma^{\rho}_{\delta\zeta}
-  \hat{u}_{\gamma , \rho , \zeta , \beta}\Gamma^{\rho}_{\delta\varepsilon}
-  \hat{u}_{\gamma , \delta ,\rho , \beta}\Gamma^{\rho}_{\zeta \varepsilon}
\, .
\end{array}
\end{equation}
Eventually the simplified equation    (\ref{eq:hatuwsimpl}) takes the form:
\begin{equation}
\label{eq:endweq}
\begin{array}{l}
\displaystyle
- c^{\alpha 3\gamma 3}w_{\gamma , 3 , 3 } +
 v^{\alpha 3\gamma 3 3 3}w_{\gamma , 3 , 3 , 3 , 3 } 
 =(c^{\alpha 3 \gamma (\beta)} + c^{\alpha (\beta) \gamma 3})
\hat{u}_{\gamma , (\beta) , 3} - \\ \\
\displaystyle
c^{\alpha\beta\gamma\delta}\Gamma^3_{\delta\beta}\hat{u}_{\gamma , 3} 
 - (c^{\alpha\beta\gamma 3} + c^{\alpha 3 \gamma\beta})
\Gamma^{\varepsilon}_{\gamma\beta}\hat{u}_{\varepsilon , 3} - 
2(v^{\alpha 3\gamma (\beta) 3 3} + \\ \\
\displaystyle
v^{\alpha (\beta) \gamma 3 3 3})
\hat{u}_{\gamma  , (\beta), 3 , 3 , 3 } 
+ 2( v^{\alpha\beta\gamma 3 3 3}\Gamma^{\rho}_{\gamma\beta} +
v^{\alpha 3\gamma\delta 3 3}\Gamma^{\rho}_{\gamma \delta})
\hat{u}_{ \rho , 3 , 3 , 3} + \\ \\
\displaystyle
(4v^{\alpha\beta\gamma\delta 3 3 } \Gamma^{3}_{\delta\beta} +
v^{\alpha\beta\gamma 3 \delta 3} \Gamma^{3}_{\delta\beta} +
v^{\alpha 3\gamma\delta 3 \beta}
\Gamma^{3}_{\delta\beta })\hat{u}_{\gamma , 3 , 3 , 3}
\, .
\end{array}
\end{equation}

Taking into account that near the body surface  the Christoffel symbols and material tensors  can be considered as constants, the equation (\ref{eq:endweq}) has a very specific form:  all of  its terms are differentiated with respect to $x^3$  at least once. Therefore, there exists a particular solution of this equation which obeys an equation with one less  derivation:
\begin{equation}
\label{eq:endweqred}
\begin{array}{l}
\displaystyle
- c^{\alpha 3\gamma 3}w_{\gamma , 3  } +
 v^{\alpha 3\gamma 3 3 3}w_{\gamma , 3 , 3 , 3  } 
 =(c^{\alpha 3 \gamma (\beta)} + c^{\alpha (\beta) \gamma 3})
\hat{u}_{\gamma , (\beta) } - \\ \\
\displaystyle
c^{\alpha\beta\gamma\delta}\Gamma^3_{\delta\beta}\hat{u}_{\gamma } 
 - (c^{\alpha\beta\gamma 3} + c^{\alpha 3 \gamma\beta})
\Gamma^{\varepsilon}_{\gamma\beta}\hat{u}_{\varepsilon } - 
2(v^{\alpha 3\gamma (\beta) 3 3} + \\ \\
\displaystyle
v^{\alpha (\beta) \gamma 3 3 3})
\hat{u}_{\gamma  , (\beta), 3 , 3  }
+ 2( v^{\alpha\beta\gamma 3 3 3}\Gamma^{\rho}_{\gamma\beta} +
v^{\alpha 3\gamma\delta 3 3}\Gamma^{\rho}_{\gamma \delta})
\hat{u}_{ \rho , 3 , 3 } + \\ \\
\displaystyle 
(4v^{\alpha\beta\gamma\delta 3 3 } \Gamma^{3}_{\delta\beta} +
v^{\alpha\beta\gamma 3 \delta 3} \Gamma^{3}_{\delta\beta} +
v^{\alpha 3\gamma\delta 3 \beta}
\Gamma^{3}_{\delta\beta })\hat{u}_{\gamma , 3 , 3 }
\, .
\end{array}
\end{equation}

Note that in the left-hand side of (\ref{eq:endweqred}) there is  exactly the expression that is needed in  (\ref{eq:sdefm8}). The general solution of the inhomogeneous differential equation is the sum of a particular solution of the inhomogeneous equation and the general solution of the homogeneous equation. But the remarkable fact is that any solution of the homogeneous equation decaying away from the surface when substituted into $ - c^{\alpha 3\gamma 3}w_{\gamma , 3  } +  v^{\alpha 3\gamma 3 3 3}w_{\gamma , 3 , 3 , 3  }$ yields zero.  So it is quite enough to consider only a particular solution, and one can simply replace $- c^{\alpha 3\gamma 3}w_{\gamma , 3  } + 
 v^{\alpha 3\gamma 3 3 3}w_{\gamma , 3 , 3 , 3  }$ in  (\ref{eq:sdefm8}) by right-hand side of (\ref{eq:endweqred}).  Having made the replacement, we obtain  $s^{\alpha}$ in the form:
\begin{equation}
\label{eq:snotready}
\begin{array}{l}
\displaystyle
s^{\alpha} =  - c^{\alpha 3\gamma (\delta)}\hat{u}_{\gamma , (\delta)} + 
c^{\alpha 3\gamma\delta}\hat{u}_{\varepsilon}
\Gamma^{\varepsilon}_{\gamma\delta} 
+(c^{\alpha 3 \gamma (\beta)} + c^{\alpha (\beta) \gamma 3})
\hat{u}_{\gamma , (\beta) } - \\ \\
\displaystyle
c^{\alpha\beta\gamma\delta}\Gamma^3_{\delta\beta}\hat{u}_{\gamma } 
 - (c^{\alpha\beta\gamma 3} + c^{\alpha 3 \gamma\beta})
\Gamma^{\varepsilon}_{\gamma\beta}\hat{u}_{\varepsilon } - 
2(v^{\alpha 3\gamma (\beta) 3 3} + \\ \\
\displaystyle
v^{\alpha (\beta) \gamma 3 3 3})
\hat{u}_{\gamma  , (\beta), 3 , 3  } 
+ 2 v^{\alpha 3\gamma (\delta) 3 3 } \hat{u}_{\gamma  , 3 , 3 , (\delta) }
+ 2 v^{\alpha 3  \gamma 3 (\beta) 3}\hat{u}_{\gamma , 3 , 3 , (\beta)} - \\ \\
\displaystyle
 h^{\alpha\beta}\hat{u}_{\beta , 3 , 3}  \, .
\end{array}
\end{equation}
Elementary transformations lead (\ref{eq:snotready}) to a rather simple form:
\begin{equation}
\label{eq:sready}
s^{\alpha} =  c^{\alpha (\beta) \gamma 3}\hat{u}_{\gamma , (\beta) } -
c^{\alpha\beta\gamma\delta}\Gamma^3_{\delta\beta}\hat{u}_{\gamma } 
 - c^{\alpha\beta\gamma 3} \Gamma^{\varepsilon}_{\gamma\beta}
 \hat{u}_{\varepsilon } - h^{\alpha\beta}\hat{u}_{\beta , 3 , 3} 
\, ,
\end{equation}
where $h^{\alpha\beta}$ is redefined as
\begin{equation}
\label{eq:hready}
h^{\alpha\beta} =  
v^{\alpha (\gamma) \beta 3 (\delta) 3 } \Gamma^3_{(\gamma)(\delta)} - 
 v^{\alpha 3\gamma 3 3 3}\Gamma^{\beta}_{\gamma 3} - 
2v^{\alpha 3 \beta\delta 3 3 } \Gamma^{3}_{\delta 3} 
- v^{\alpha\varepsilon\beta 3 \delta 3} \Gamma^{3}_{\delta\varepsilon}
 \, .
\end{equation}
This equations  together  with (\ref{eq:bond3m2})  completely  solves the problem considered in this section.

Remember that in used coordinate system unit vector normal to  the body surface $n_{\alpha}$ has the only component $n_3=1/\sqrt{g^{33}}$. So that with respect to the classical part of strain, $\left. c^{\alpha 3\gamma\delta}\tilde{u}_{\gamma;\delta}\right|_S$ actually defines the external force density $\sigma^{\alpha\beta}n_{\beta}$ on the body surface, one should only multiply it by $1/\sqrt{g^{33}}$. This is why differential equations (\ref{eq:tildeeq}) appended by  (\ref{eq:bond3m2}) correspond to a standard problem of classical theory of elasticity for a body under external forces on its surface. Methods of solution of such problems are well-known and do not require special consideration.

It should be emphasized that physically there are no external forces on the surface. Forces mentioned above are purely formal in nature and describe the interaction of non-classical and classical parts of deformation.

\section{Comparison of exact and approximate solutions for a homogeneously polarized ball}

It worth to find flexoelectric deformations of homogeneous polarized ball by approximate method described in previous section and compare this deformations with ones known from exact solution. This is a test of approximate method at least.

For isotropic material the matrices $c^{\alpha 3 \gamma 3}$ and $v^{\alpha 3 \gamma 3 3 3}$ are diagonal.  So that in order to find $\hat{u}_{\gamma}$ one does not even need to solve the generalized eigenvalue problem  (\ref{eq:eigen}), the result is obtained at once:
\begin{equation}
\left\{
\begin{array}{l}
\displaystyle
\hat{u}_r= \frac{f_{12}+2f_{44}}{2(c_{12}+2c_{44})} P \cos \theta
e^{\lambda_r R(\xi - 1)} \, ,\\ \\
\displaystyle
\hat{u}_{\theta}= - R\frac{f_{44}}{2c_{44}} P \sin \theta 
e^{\lambda_{\theta} R(\xi - 1)} \, ,\\ \\
\hat{u}_{\psi}=0 \, ,
\end{array}
\right.
\end{equation}
where
\begin{equation}
\lambda_{\theta}= \sqrt{\frac{c_{44}}{v_1+2v_2}} \, ,
\end{equation}
\begin{equation}
\lambda_r=\sqrt{\frac{c_{12}+2c_{44}}{9v_1+6v_2}} \, .
\end{equation}
For comparison with the exact solution it is useful to present $\hat{u}_{\gamma}$ in terms of the functions $\hat{f}_1$ and $\hat{f}_2$:
\begin{equation}
\hat{f}_1 = \frac{\hat{u}_r}{\cos \theta} =
\frac{f_{12}+2f_{44}}{2(c_{12}+2c_{44})} P e^{\lambda_r R(\xi - 1)} \, ,
\end{equation}
\begin{equation}
\hat{f}_2 = - \frac{\hat{u}_{\theta}}{r\sin \theta} =
\frac{1}{\xi}\cdot\frac{f_{44}}{2c_{44}} P  e^{\lambda_{\theta} R(\xi - 1)} \, .
\end{equation}

Next step is to find $\tilde{\sigma}^{\alpha 3}n_3=c^{\alpha 3\gamma\delta}\tilde{u}_{\gamma;\delta}n_3$ on the ball surface. To do this,  one should just use the equations of the previous section. It turns out
\begin{equation}
\label{eq:tildes33ball}
\left.\tilde{\sigma}^{33}n_3\right|_S= \frac{P \cos \theta}{R}\left[
\frac{2c_{44}(f_{12}+2f_{44})}{c_{12}+2c_{44}} - 2f_{44}
\right] \, ,
\end{equation}
\vspace{0.5cm}
\begin{equation}
\label{eq:tildes23ball}
\left.\tilde{\sigma}^{23}n_3\right|_S= \frac{P \sin \theta}{R^2}\left[
\frac{f_{12}}{2} - \frac{c_{12}(f_{12}+2f_{44})}{2(c_{12}+2c_{44})}
\right] \, .
\end{equation}
Certainly $\tilde{\sigma}^{13}=0$. This is obviously from symmetry but can be also obtained by direct calculations.

To solve the differential equations (\ref{eq:tildeeq})  actually  is not necessary here. It is  clear that in the terms of functions $\tilde{f}_1=\tilde{u}_r/\cos\theta$ and \mbox{$\tilde{f}_2=-\tilde{u}_{\theta}/(r\sin\theta)$} the solution  is proportional to the previously defined function ${\cal B}_1$. Eventually the validity of this statement can be verified by direct substitution. So that one just needs to find the coefficient of proportionality, it is easily done by using (\ref{eq:tildes33ball}) or  (\ref{eq:tildes23ball}). One  can use any of these conditions, they both give the same result:
\begin{equation}
\tilde{f}_1=\frac
{P(c_{12}f_{44}-c_{44}f_{12})(c_{12}-c_{44})}
{c_{44}(c_{12}+2c_{44})(3c_{12}+2c_{44})} \,\xi^2 \, ,
\end{equation}
\vspace{0.5cm}
\begin{equation}
\tilde{f}_2=\frac
{P(c_{12}f_{44}-c_{44}f_{12})(2c_{12}+3c_{44})}
{c_{44}(c_{12}+2c_{44})(3c_{12}+2c_{44})}\,\xi^2 \, .
\end{equation}

It is also useful to derive  the expression for the classical part of the elastic displacement in Cartesian components as a function of the Cartesian coordinates. Direct conversions yield
\begin{equation}
\left\{
\begin{array}{l}
\displaystyle
\tilde{u}_z=a_1 z^2 + a_2(x^2+y^2) \, , \\
\displaystyle 
\tilde{u}_y=(a_1-a_2) yz \, , \\
\displaystyle
\tilde{u}_x=(a_1-a_2) xz \, ,
\end{array}
\right.
\end{equation}
where
\begin{equation}
a_1=\frac
{P(c_{12}f_{44}-c_{44}f_{12})(c_{12}-c_{44})}
{R^2c_{44}(c_{12}+2c_{44})(3c_{12}+2c_{44})} \, ,
\end{equation}
\vspace{0.5cm}
\begin{equation}
a_2=\frac
{P(c_{12}f_{44}-c_{44}f_{12})(2c_{12}+3c_{44})}
{R^2c_{44}(c_{12}+2c_{44})(3c_{12}+2c_{44})} \, .
\end{equation}

\begin{figure}[h]
\begin{center}
\includegraphics[width=10cm,keepaspectratio]{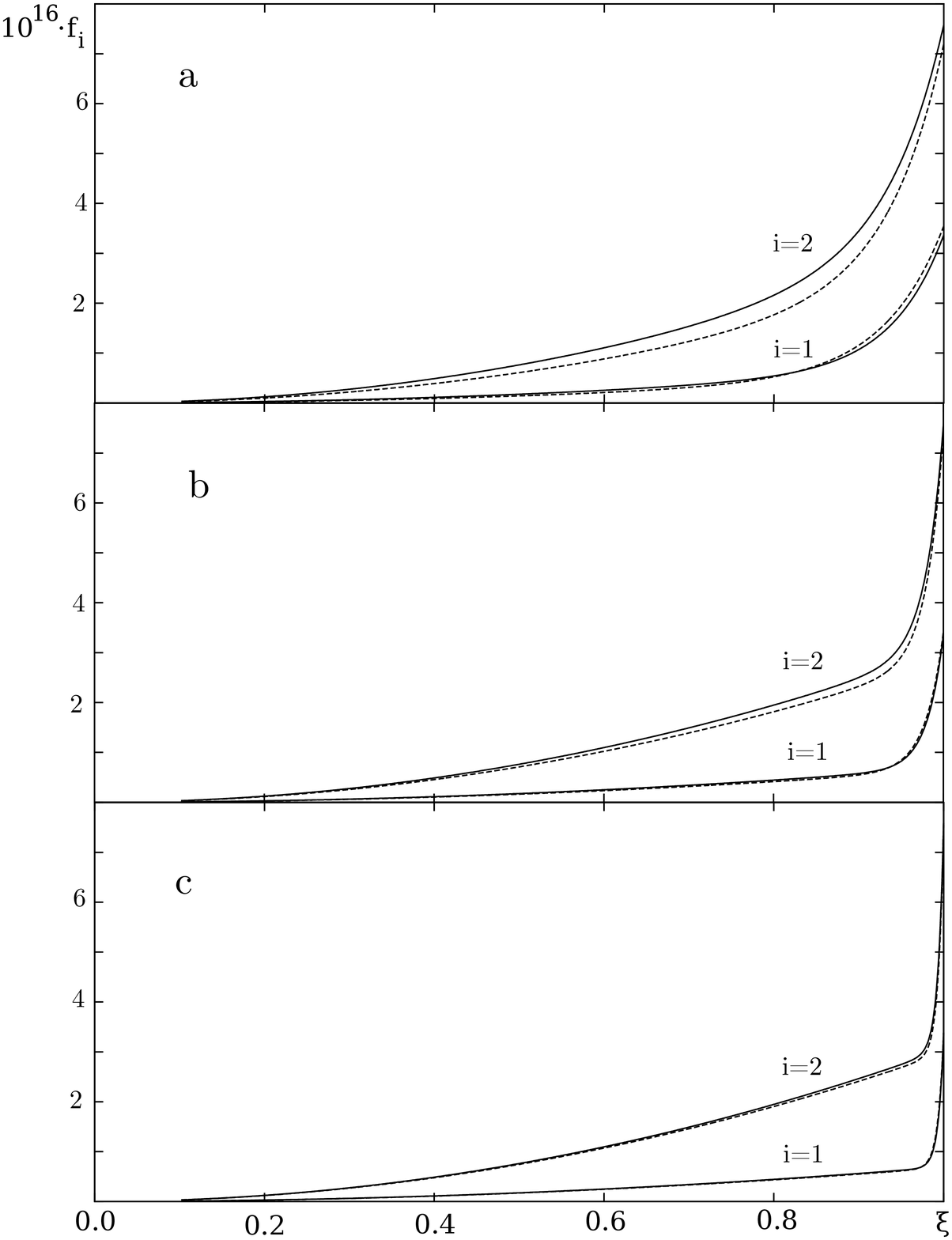}
\end{center}
%
\caption{Comparison of exact and approximate solutions for a ball. $R=1\cdot 10^{-5}\,$, $P=1\,$,  $c_{44}=1.1\cdot 10^{12}\,$, $c_{12}=3.4\cdot 10^{12}\, $, $f_{44}=f_{12}=1\cdot 10^{-3}\, $. $a$ -- for $v_1=2.0 \cdot 10^{-1}\,$,  $v_2=1.0 \cdot 10^{-1}\,$; $b$ -- for  $v_1=2.0 \cdot 10^{-2}\,$,  $v_2=1.0 \cdot 10^{-2}\,$; $c$ -- for  $v_1=2.0 \cdot 10^{-3}\,$,  $v_2=1.0 \cdot 10^{-3}\,$. Solid line is approximate solution, dashed line is exact solution. }
\label{fig:ballcompare}
\end{figure}

Now we can find $f_i=\hat{f}_i+\tilde{f}_i$ and compare it with the results of exact calculations. This comparison is shown in Fig.\ref{fig:ballcompare}. This figure shows a good agreement between the approximate and the exact solution, the smaller  higher elastic moduli match those better. It should also be emphasized that the latest version, when the difference is badly distinguishable, corresponds to most physically reasonable values of higher elastic moduli . Thus, the approximate method works very well at least for a homogeneously polarized ball with the given parameters.

%

\clearpage

\section{Flexoelectric bending of homogeneously polarized  circular rod}

In this section we apply the above-described method   to find the bending of a homogeneously polarized circular rod of an isotropic material \cite{bib:YuPSS14}. Such a rod may be polarized in different directions. However, due to the fact that the flexoelectric effect is linear, we can restrict ourselves to the case of longitudinal and transverse polarization only. The response of the rod to the polarization of an arbitrary direction  obviously is a superposition of responses to the longitudinal and transverse polarization. Besides, axial symmetry allows us not distinguish different directions of the transverse polarization.

As stated above, the classical part of strain is determined by the formal forces on the body surface. The surface of the rod consists of a cylindrical surface and end surfaces. First we consider the cylindrical surface of the rod. Here it is natural to use a cylindrical coordinate system. Let  cylinder be located along the longitudinal  axis corresponded to coordinate $x$. This coordinate is also denoted as $x^1$. The other two curvilinear coordinates are the angular $x^2=\phi$ and  radial $x^3=r$. Angle $\phi$ is measured in respect to the direction of the Cartesian $z$-axis. Equation of the surface is $x^3=R$ where $R$ is rod radius. So that this  coordinate system belongs to the class that is needed in  accordance with the  method of calculations.

For such coordinates  simple geometrical reasoning  lead to the following equations defining the relationship between  Cartesian and curvilinear coordinates:
\begin{equation}
\left\{
\begin{array}{l}
\displaystyle
x= x^1  \, , \\ \\
\displaystyle
y=r \sin \phi = x^3 \sin x^2 \, , \\ \\

\displaystyle
z=r \cos \phi = x^3 \cos x^2 \, .
\end{array}
\right.
\end{equation}
By means of simple differentiation it  is easy to find the metric tensor, its determinants and the Christoffel symbols:
\begin{equation}
\left\{
\begin{array}{l}
\displaystyle
g_{11}=g_{xx}= 1 \, , \quad
g_{22}=g_{\phi\phi}= r^2 \, , \quad
g_{33}=g_{rr}= 1 \, , \quad
g=r^2 \, , \\ \\
\displaystyle
\Gamma^r_{\phi\phi} = -r \, , \quad
\Gamma^{\phi}_{ r \phi} = \Gamma^{\phi}_{\phi r} =  \frac{1}{r} \, .
\end{array}
\right.
\end{equation}
The remaining components are zero. Since the metric tensor $g_{\alpha\beta}$ is diagonal to find the inverse  metric tensor $g^{\alpha\beta}$ is easy: one just need to take the reciprocals of the diagonal components. Note also that the vector normal to  surface has the only component $n_r=1$.

By making perfectly straightforward  calculations  similar to the case of a ball described above, the following expression can be obtained for the non-classical part of the elastic displacement $\hat{u}_{\alpha}$ near the cylindrical surface. If the polarization $P$ is longitudinal along the $x$-axis then the  expression is as follows:
\begin{equation}
\left\{
\begin{array}{l}
\displaystyle
\hat{u}_x=\hat{u}_1= P \frac{f_{44}}{2c_{44}}  \, e^{\lambda_xR(\xi-1)} \, , \\ \\
\displaystyle
\hat{u}_{\phi} =\hat{u}_2= 0 \, , \\ \\
\displaystyle
\hat{u}_r = \hat{u}_3= 0 \, .
\end{array}
\right.
\end{equation}
For  transverse polarization along $z$-axis $\hat{u}_{\alpha}$ is determined by the following equations:
\begin{equation}
\left\{
\begin{array}{l}
\displaystyle
\hat{u}_x=0 \, , \\ \\
\displaystyle
\hat{u}_{\phi} = - PR \frac{f_{44}}{2c_{44}}\sin\phi \, e^{\lambda_{\phi}R(\xi-1)} \, ,\\ \\
\displaystyle
\hat{u}_r =  P\,\frac{f_{12}+2f_{44}}{2(c_{12}+2c_{44})}\cos\phi \, e^{\lambda_rR(\xi-1)} \, .
\end{array}
\right.
\end{equation}
Here $\xi=r/R$ is dimensionless radial coordinate,  
\begin{equation}
\lambda_x=\lambda_{\phi}=\sqrt{\frac{c_{44}}{v_1+2v_2}} \, ,
\end{equation}
\begin{equation}
\lambda_r=\sqrt{\frac{c_{12}+2c_{44}}{9v_1+6v_2}} \, .
\end{equation}

Further,  direct application of the equations of  section \ref{sec:aprox} gives the following boundary conditions for the classical part of the displacement. For transverse polarization they are
\begin{equation}
\label{eq:rodbc1}
\left \{
\begin{array}{l}
\displaystyle
\left.\tilde{\sigma}^{13}n_3\right|_S = 0 \, , \\ \\
\displaystyle
\left.\tilde{\sigma}^{23}n_3\right|_S = - \frac{P\sin\phi}{R^2} \cdot \frac{1}{2}\left[
\frac{c_{12}(f_{12}+2f_{44})}{c_{12}+2c_{44}} -f_{12}
\right] \, , \\ \\
\displaystyle
\left.\tilde{\sigma}^{33}n_3\right|_S = \frac{P\cos\phi}{R}\left[
\frac{c_{44}(f_{12}+2f_{44})}{c_{12}+2c_{44}} - f_{44}
\right] \, .
\end{array}
\right.
\end{equation}
If polarization is longitudinal then all $\tilde{\sigma}^{\alpha 3}$ are zero.

Boundary conditions (\ref{eq:rodbc1}) can be also expressed  in terms of the Cartesian  tensor component. Elementary transformations  yield:
\begin{equation}
\label{eq:foncylcart}
\left\{
\begin{array}{l}
\displaystyle
\left.\tilde{\sigma}_{xi}n_i\right|_S = 0 \, , \\ \\
\displaystyle 
\left.\tilde{\sigma}_{yi}n_i\right|_S = \frac{P(c_{44}f_{12}-c_{12}f_{44})}{R(c_{12}+2c_{44})}
\cdot  \sin 2 \phi  \, , \\ \\
\displaystyle
\left.\tilde{\sigma}_{zi}n_i\right|_S = \frac{P(c_{44}f_{12}-c_{12}f_{44})}{R(c_{12}+2c_{44})}
\cdot \cos 2\phi  \, .
\end{array} \right.
\end{equation}
In this form of boundary conditions it is immediately clear that the total force acting on the cylindrical surface is equal to zero, the integration over the angle turns these expressions to zero. It also can be easily  calculated that the total bending moment of these forces vanishes while  integrating over the angle $\phi\,$.

Thus, not only the sum  of formal surface forces, but also the sum of the bending moments of these forces, acting on a small part of the cylindrical surface length, are equal to zero. This means that the bending moment does not change along  the homogeneously polarized  rod. We emphasize that even so, the bending moment appears, but entirely by the boundary effects at the ends of the rod. So that in the case of homogeneously polarized rod the problem  is reduced to the standard problem of the classical theory of elasticity \cite{bib:LL7},  i.e. to the determination of rod  bending  under the  bending moments applied to its ends. 

Certainly the surface forces (\ref{eq:rodbc1}) slightly deform cross-section of the rod. Since the main effect is the bending of the rod, we do not discuss this deformation. Note only that qualitatively it  is similar to the deformation of the ball in the meridian cross-section.

To calculate the bending moment appeared at the ends of the rod,  we need to specify the shape of these ends. We emphasize that we can not limit the rod by planes because this gives  sharp edges while the theory requires  smooth surface. We should  smooth out these sharp edges, say by a quarter of toroidal surface,  or assume that the rod terminates by halves of the ball. Calculations were made for both these variants, and it was obtained the same result. Omitting the details of these rather simple calculations, we note only that in the case of spherical ends one  can use the results of the previous section. For edges  slightly smoothed by the toroidal surface one can approximately describe a small part  of this toroidal surface as cylindrical one and use the equations given above in this section.  Eventually it turns out that the bending moment  is
\begin{equation}
\label{eq:myrod}
M_y =  \frac{P\pi R^2(c_{44}f_{12}-c_{12}f_{44})}{c_{12}+2c_{44}} 
 \, .
\end{equation}
The other components are equal to zero. It is  implied here  that the rod is transversely polarized along  $z$-axis  and the equation is written for  the end  in the positive direction of  $x$-axis. For the other  end of the rod $M_y$  has the opposite sign. 

As is already clear from symmetry,  for  the longitudinal polarized rod there is no bending moment  at all, this can be also proved by direct calculations. Such calculations also show that the total force acting on each end of the rod is  zero as for  longitudinal and for  transverse polarization.

Equation (\ref{eq:myrod}) actually solves the problem of flexoelectric bending of the homogeneously polarized rod. One should  only substitute this bending moment into  standard equations from textbooks (see \cite{bib:LL7} for instance). 

\section{Flexoelectric bending of a homogeneously polarized  circular plate}

In this section we apply the above-described method of approximate calculation  to find  the flexoelectric bending of a thin circular plate with a radius $ R $ and thickness $ h $, uniformly polarized normal to its plane \cite{bib:YuPSS14}. Material of the plate is assumed  isotropic. For definiteness, we  assume that the average surface of the plate lies in the coordinate plane $ OXY $.

Calculations for the plate are generally similar to those made in the previous section for the rod. However, most part of the plate surface is flat, so that one can immediately conclude, without any calculations,  that the formal forces on this part of the surface are zero. As in the case of the rod, bending of the plate is determined by edge effects which  are discussed below.

Thus, for  most part of the plate, except for the edges,  the classical part of the plate deformation is determined by standard equations  without surface loading. It is well known \cite{bib:LL7} that under such conditions the displacements of the average surface of the plate $\zeta(x,y)$  obey  two-dimensional differential equation
\begin{equation}
\label{eq:bilaplace}
\Delta\Delta\zeta =0 \, ,
\end{equation}
where $\Delta$ is  two-dimensional Laplace operator. The components of the strain tensor can be expressed in terms of $\zeta$ as follows:
\begin{equation}
\label{eq:uab}
\begin{array}{l}
\displaystyle
u_{xx}=-z\zeta_{,x,x} \quad ; \quad
u_{yy}=-z\zeta_{,y,y} \quad ; \quad
u_{xy}=-z\zeta_{,x,y} \quad ; \\
\displaystyle
u_{xz}=u_{yz}=0 \quad ; \quad
u_{zz}=z\,\,\frac{c_{12}}{c_{12}+2c_{44}}\,\, (\zeta_{,x,x}+\zeta_{,y,y}) \quad .
\end{array} 
\end{equation}

Using polar coordinates  it is easy to find an axially symmetric, regular at origin,  solution of the equation (\ref{eq:bilaplace}):
\begin{equation}
\label{eq:zetar}
\zeta(x,y) = -\frac{Gr^2}{2} =-\frac{G}{2}(x^2+y^2) \, .
\end{equation}
This solution contains a single  integration constant $G$ which is nothing but the curvature of the plate. 

To find  integration constant $G$ one needs first calculate the components of the strain tensor by means of  (\ref{eq:uab}) and second calculate the components of the strain tensor  by means of standard equations $\sigma_{ij}=c_{ijkl}u_{ij}\,$. Thereafter  it becomes obvious that  there is the linear density of the bending moment $M$ on the boundary  of the plate. For instance, at the point of intersection of the plate boundary with the coordinate axis $OX$ this bending moment density  is
\begin{equation}
\label{eq:msigma}
M=\int\limits_{-h/2}^{+h/2} \sigma_{xx}(z) z dz 
= \frac{h^3}{6}\,\, G\,\,\frac{c_{44}(3c_{12}+2c_{44})}{c_{12}+2c_{44}} 
\, .
\end{equation}
Due to axial symmetry  $M$ is the same at all other points of boundary.

On the other hand  $M$  can be calculated in terms of formal forces arising due to flexoelectricity on curved surfaces on the plate boundary. As in the case of the rod,  it is necessary to smooth out sharp edges using one-quarter of toroidal surface (on each sharp edge) or one-half of the toroidal surface for the whole boundary. A small part of a toroidal surface can be considered as a cylindrical one,  so that we can apply the equations  (\ref{eq:foncylcart}) (coordinate system should be rotated). For both variants of sharp edges smoothing it turns out
\begin{equation}
M= \frac{Ph(c_{44}f_{12}-c_{12}f_{44})}{c_{12}+2c_{44}} \, .
\end{equation}
It only remains to equate the two expressions for $M$ and express the curvature of the plate  in terms of polarization $P$:
\begin{equation}
\label{eq:flexg}
G=\frac{6P(c_{44}f_{12}-c_{12}f_{44})}{h^2c_{44}(3c_{12}+2c_{44})} \, .
\end{equation}

It is interesting to compare the equation (\ref{eq:flexg}) with the result obtained in \cite{bib:TagYu12} by direct minimization of the plate energy. Elementary transformations of the equations from \cite{bib:TagYu12} yield the equation different from (\ref{eq:flexg}) only in that there is 12 instead 6 in the numerator. However, it should be kept in mind that in \cite{bib:TagYu12} it is considered the case when in a thin  layer near the plate surface  polarization falls to zero. If, in accordance with calculations presented here,  we modify the calculations \cite{bib:TagYu12} to the case, when the polarization is strictly homogeneous,  then  it appears  exactly the equation (\ref{eq:flexg}).  

A similar dependence on the boundary conditions imposed on the polarization also holds for direct flexoelectric effect, it  is discussed in detail in another paper \cite{bib:YTD}.

\section{Conclusion}

Above we have considered  the continuum theory of the flexoelectric effect in the finite-size bodies. In the part of description of the converse flexoelectric effect,  this theory turns out to be quite complicated. This is due to the fact that the independent variables are the elastic displacement. Flexoelectric part of the thermodynamic potential depends on the second spatial derivatives of the elastic displacement. Moreover  in the case of flexoelectricity the elastic energy  should be considered taking into account  its dependence on the second spatial derivatives of elastic displacement (elastic spatial dispersion), or the theory is self-contradictory in the general case. All of this leads to several theoretical problems.

The first theoretical problem is that   derivation of the boundary conditions for the differential equations of elastic equilibrium becomes non-trivial in such a case. While the volume  integration by parts, which is necessary  to express the variation of the thermodynamic potential in terms of  independent variates, its appear the surface integrals containing the gradients of these independent variates. Needed transformation of these surface integrals require special mathematical tools. For this purpose it was offered to  use additional surface curvilinear coordinate system or  to solve the problem in curvilinear coordinates of the general form from the beginning. From a practical point of view the second approach is more convenient although it has been proved that this two approaches are mathematically equivalent. It should also be noted that its appear that the curvature of the boundary surface plays an important role in the  boundary conditions obtained in both approaches.

Solving the problem of elastic boundary conditions derivation, we get the opportunity to solve, in principle, any boundary problems needed to describe flexoelectricity in finite-size bodies. However, the corresponding boundary problems are extremely complex and difficult to solve. Such a boundary  problem has been solved for a homogeneously  polarized ball but for more complex geometries to make a similar is unrealistic. Even for the ball the solution  is extremely cumbersome. Therefore  the development of approximate methods for solving such boundary problems is desired. This is  the second theoretical problem.

The problem of developing a method for the approximate solution of the corresponding boundary problems were also solved and the solution is described above. It turns out that in the framework of such problems elastic displacements can be approximated   as the sum of two parts. The first part  was called non-classical, it is concentrated near the surface of the body and decays exponentially inside the body. To find this part it is necessary to solve fairly simple one-dimensional equations. The second, classical part is determined by the equations of the classical theory of elasticity.   It turns out  that the boundary conditions for this  equations  have a standard classical form of the boundary conditions for the body under the  external forces on the surface. Thus, the finding of the classical part is reduced to the standard classical problem which does not require a separate discussion. It should be noted that these forces on the surface are  formal. Physically there is no  forces on surface,   formal forces  describe the interaction between classical and non-classical part of the elastic displacement. So that these formal forces are expressed in terms of non-classical part of the displacement. The corresponding equations are given.

It is important to note that for homogeneous polarization the formal forces, describing the influence  of the non-classical part of displacement to the  classical one, appear only on the curved parts of the surface of the body. Thus, the above statement that  curvature of the surface plays an important role gets a clear physical meaning. Note that the body can be, say, a polyhedron. In this case the formal forces appear only on the edges which can be treated as a limiting case of a curved surface. While  calculating one should slightly smooth these edges and tend to zero the radius of smoothing at the end of the calculations.  The similar approach  is applied above for the particular case of the rod and thin plate  which also has the  sharp edges.

Thus,  in framework of the theory of continuum the description flexoelectricity in finite-size bodies  actually requires special  theory. This  theory is described above in detail. The application of this theory  is illustrated in particular problems of  calculation of the bending of  homogeneously polarized rod and a thin plate.

\section*{Acknowledgements}

A.K.~Tagantsev is acknowledged for reading the manuscript. 


\end{document}